\documentclass[aps,superscriptaddress,showpacs]{revtex4}            

\newcommand{\st}{$\sigma_{tot}$}
\newcommand{\ro}{$\rho$}

\usepackage{epsfig}

\begin{document}

\title{\bf Analytic models and forward scattering from accelerator 
to 
cosmic-ray energies}  

\date{\today}

\author{R. F. \'Avila} 
\email{rfa@ifi.unicamp.br}
\affiliation{Instituto de Matem\'atica, Estat\'{\i}stica e 
Computa\c c\~ao
Cient\'{\i}fica \\
Universidade Estadual de Campinas,
13083-970 Campinas, SP, Brazil}
\author{E. G. S. Luna}
\email{luna@ifi.unicamp.br}
\affiliation{Instituto de F\'{\i}sica Gleb Wataghin, \\
Universidade Estadual de Campinas,
13083-970 Campinas, SP, Brazil}
\author{M. J. Menon}
\email{menon@ifi.unicamp.br}
\affiliation{Instituto de F\'{\i}sica Gleb Wataghin, \\
Universidade Estadual de Campinas,
13083-970 Campinas, SP, Brazil}

\begin{abstract}
Analytic models for hadron-hadron scattering are characterized by  
simple analytical parametrizations for the forward amplitudes and 
the 
use of dispersion 
relation techniques 
to study the total cross section $\sigma_{tot}$ and the $\rho$ 
parameter
(the ratio between the real and imaginary parts of the forward amplitude). 
In this paper we investigate simultaneously four aspects  
related to the application of the model to $pp$ and 
$\bar{p}p$ scattering, from accelerator to cosmic-ray energies: (1) 
the effect of  
different estimations for  $\sigma_{tot}$ from cosmic-ray experiments; 
(2) the differences between individual and global (simultaneous) fits to 
$\sigma_{tot}$ 
and $\rho$; (3) the role of the subtraction constant in the dispersion 
relations;
(4) the effect of distinct asymptotic inputs from different analytic models. 
This is done
by using as a framework  
the single Pomeron and the maximal Odderon parametrizations for the
total cross section. Our main conclusions are the following: 
(1) Despite the small influence from different cosmic-ray estimations, the
results allow us to extract an upper bound for the soft Pomeron 
intercept: $1 + \epsilon = 1.094$;
(2) although global fits present good statistical results, in general, this
procedure constraints the rise of  $\sigma_{tot}$; 
(3) the subtraction constant as a free
parameter affects the fit results at both low and high energies; 
(4) independently of the 
cosmic-ray information used and the subtraction constant, global fits with 
the Odderon 
parametrization predict that, above $\sqrt s \approx 70$ GeV, $\rho_{pp}(s)$ 
becomes greater 
than $\rho_{\bar{p}p}(s)$, and this result is in complete
agreement with all the data presently available. In particular, we infer
$\rho_{pp} = 0.134\ \pm \ 0.005$ at $\sqrt s = 200$ GeV and $0.151\ \pm 
\ 0.007$ at $500$ GeV (BNL RHIC energies). 
A detailed discussion of the procedures used and all the results obtained is 
also presented.
\end{abstract}


\maketitle

\section{Introduction}

High-energy soft processes are presently a topical problem 
in high-energy physics, mainly because they are essentially
a nonperturbative phenomenon \cite{dok}. 
Despite important results at the interface between soft and hard physics
and recent progress in nonperturbative QCD,  \textit{elastic scattering}, 
the
simplest soft process, cannot be described in a pure QCD framework, but
only in a phenomenological context.

In this area, a variety of models, such as, for example, Regge, Diffraction, 
QCD-inspired models, and others \cite{matthiae},
has survived due to a solid theoretical
basis and  efficiency in describing physical quantities.
What is presently well established  and accepted concerns some general
principles, limits, bounds, and theorems, earlier deduced from the
Mandelstam representation, potential scattering, and also from axiomatic 
field theories 
\cite{eden,het}. In this context, analyticity, unitarity and crossing 
play  central roles and are
also the framework of several models referred to above. Among them, the so
called analytic models are characterized by parametrizations of the
hadronic amplitude through general analytic functions that strictly obey the
formal principles and theorems.
Specifically, the aim is to investigate two fundamental physical
quantities that characterize the forward elastic scattering at high energies,
namely, the total cross
section $\sigma_{tot}$ and the $\rho$ parameter (the ratio of the real to
the imaginary part of the forward elastic scattering amplitude). 
In terms of the scattering amplitude $F$ they may be given by

\begin{eqnarray} 
\sigma_{tot}(s) = \frac{\textrm{Im}\ F(s,t=0)}{s}, 
\qquad
\rho(s) = \frac{\textrm{Re}\ F(s,t=0)}{\textrm{Im}\ F(s,t=0)}, 
\end{eqnarray} 
where $t$ is the four-momentum transfer squared and $s$ the center-of-mass
energy squared. These expressions, in terms of 
real and imaginary parts of the amplitude, obviously 
suggest dispersion relations as a suitable formal framework for 
investigations.
For particle-particle and particle-antiparticle interactions,
the addition of crossing symmetry extends and completes the analytical approach.
It is expected that such a general formalism, avoiding details of the 
interaction or dynamics, could be a suitable tool in the search for adequate 
calculational schemes in nonperturbative QCD.

The analytic models have a long history and important results have
been obtained through both \textit{integral} relations \cite{idr}, for 
example 
\cite{amaldi,bc,ua4/2,bertini}, and \textit{derivative} 
(\textit{analyticity}) relations
\cite{ddr,bks,kn}, for example \cite{kn,kvw,cudelletall}.
Recently, much effort has been concentrated in the COMPETE Collaboration
(COmputerized Models, Parameter Evaluation for Theory and Experiment),
which joined the COMPAS group and others specialists also with
outstanding contributions in the area.
These authors have investigated a large class of analytic models, 
through several statistical indicators that complement the usual $\chi^2$ and
and $C.L.$ criteria. One of the main results is the universality of the
$B\ln^2 s/s_0$ increase of the total cross section for all the collisions
considered \cite{cudelletall,competework}.

Despite all the experience accumulated and the detailed analyses that have 
been developed,
we understand that three aspects yet need some investigation and this is the
main purpose of this work. These aspects are based on the following
observations. 

\begin{description}
\item[1.] Beyond the energy region of the accelerators, 
experimental information on $pp$ total cross sections exists from cosmic-ray
experiments. Despite the model dependence involved, the large error bars in the 
numerical results, and also the existence of discrepant values, some of these 
results are usually displayed as a support for several model predictions. 
However, this set of experimental information is not usually taken into account 
in 
explicitly \textit{quantitative analyses} that could provide, for example,  
bounds
for the increase of $\sigma_{tot}$ and $\rho$ (and, consequently,  
the intercept of exchanged trajectories) or for the differences between $pp$ 
and $\bar{p}p$
total cross sections (for exceptions, see \cite{bjp01,alm}). Moreover, as will 
be discussed in
some detail, several results obtained by different authors, through different 
approaches, indicate 
nearly the same increase of $\sigma_{tot}(s)$ and this increase is faster than 
usually believed or accepted.
\item[2.] In general, the fits are performed with the full hadronic amplitude, 
that is,
simultaneous fits to $\sigma_{tot}$ and $\rho$. Although some authors correctly
claim that this procedure maximizes the number of data points, it must first be 
recalled that 
important results have been obtained through fits to only the \st data; a 
classical example is
the approach by Donnachie and Landshoff \cite{dl}. Moreover, as will be 
discussed, 
in the bulk of 
experimental data available, \st and \ro have different
status as physical quantities, since \ro is estimated either by extrapolations 
through
dispersion relations (and fits to \st data), or as a fit parameter to the 
differential 
cross
section in the region of Coulomb-nuclear interference, and this is a delicate 
problem.
Finally, since from the previous observation we 
shall also be interested in the high-energy cosmic-ray results, which concern 
only \st
(and not \ro), global fits may constrain the possible increase of this quantity. 
\item[3.] The connection between $\sigma_{tot}$ and $\rho$ through
\textit{standard dispersion relations} demands one subtraction \cite{idr}. 
Although the
subtraction constant works as a fit parameter in traditional 
analysis  \cite{amaldi,bc,ua4/2,bertini}, it does not appear in the 
approach by the COMPETE Collaboration, since the derivative relation or 
prescription 
used does not
involve subtraction constant. Once the analytical approach demands fits to 
experimental data
and the free parameters involved are all \textit{correlated}, it is expected 
that the presence or
not of the subtraction constant may lead to different results.
\end{description}

The aim of this work is to investigate the above three observations and their 
inter-connections in a
quantitative way. To this end, we shall analyze only $pp$ and $\bar{p}p$ 
elastic scattering,
since for particle and antiparticle interactions they correspond to the 
highest energy
interval with available data and are the only set including the cosmic-ray 
information on total cross
sections ($pp$ scattering). We shall  use as framework two well known analytic 
models characterized
by \textit{distinct asymptotic inputs}, the Pomeron-Reggeon model by Donnachie 
and Landshoff \cite{dl}
and the maximal Odderon model in the form discussed by Kang and Nicolescu 
\cite{kn}. For the connections 
between
real and imaginary parts of the forward amplitude we shall use 
\textit{derivative} (\textit{analyticity})
relations with \textit{one subtraction}. 
Observation 1 is treated by the selection of two different ensembles of 
data from cosmic-ray
results, observation 2 through individual and simultaneous fits to \st and 
\ro, and observation 
3 by treating the subtraction constant as a free fit parameter or assuming 
that its value is zero.

The manuscript is organized as follows. In Sec. II we present the experimental 
data
to be analyzed and the criteria for the selection of two ensembles of data. 
In Sec. III we review the main formulas
in the analytical approach, including the dispersion relations, the models
to be used and some high-energy theorems. The fits and results for all the 
cases
considered are presented in Sec. IV and the conclusions and final remarks 
in Sec. V.

\section{Experimental information and ensembles}

In this section we review the experimental information that  constitutes our 
data
base (accelerator and cosmic-ray regions), together with discussions concerning 
the
differences in the determination of \st and \ro, as well as the model 
dependences
and discrepancies related to \textit{all} the cosmic-ray information 
presently
available. The criteria for selecting two ensembles of data are also 
presented
and justified in detail.

\subsection{Accelerator data}

In the case of accelerator experiments, data on $\sigma_{tot}$ 
from $pp$ and $\overline{p}p$  scattering and extracted values for the
$\rho$ parameter have been accumulated for a long time.
Presently, experimental information extends up to $62.5$ GeV and $1.8$ TeV 
for $pp$ and $\overline{p}p$ scattering, respectively. The database, 
analyzed and
compiled by the Particle Data Group (PDG), has become a standard reference
and the corresponding readable files are available \cite{pdg}. 
Since recent analysis has shown that general fits to these data are stable 
for $\sqrt s$ above
$\sim$ 9 GeV \cite{cudelletall}, we shall use here the sets for
energies above $10$ GeV; in our analysis
the statistic and systematic errors are added linearly.

At this point, we briefly recall some differences in the determination of
\st and \ro \cite{matthiae}, which suggest that they do not have the same 
status as
physical quantities; this, in turn, corroborates our motivation to
investigate global and individual fits separately.

First, one of the methods that was used to extract \st at the ISR did not 
depend on the
\ro value, but only on the machine luminosity and the rates of elastic and 
inelastic
interactions. Therefore in these determinations \st and \ro are independent 
quantities.
Other methods demand the determination of \ro and the elastic scattering 
rate at small
momentum transfer, extrapolated to the forward direction (slope parameter). 
In these cases,
the quantities to be determined correspond to $\sigma_{tot}^{2}(1 + \rho^2)$ 
(luminosity dependent) or $\sigma_{tot}(1 + \rho^2)$ (luminosity independent) 
\cite{matthiae}.
In both cases, since it is known that $\rho < 0.14$, \st may be obtained with 
reasonable
accuracy even with a rough estimation of the \ro parameter.
Specifically, some authors use the \ro value
extrapolated from fits to \st and  dispersion relations; in the other 
procedure,
\ro is determined from fits to the differential cross section data in the 
region of
Coulomb-nuclear interference. In this case the determination is model dependent 
and it
is interesting to note that the procedure, in the last instance, demands  
knowledge of 
how the 
hadronic exchanged object interacts,
which is exactly what is looked for. This point seems clear in the 
Donnachie-Landshoff 
approach, since the authors do not use the \ro data as input.
We understand that all these facts reinforce the differences in the 
determination of \ro and \st,
putting some limits on the interpretation of \st and \ro as physical 
quantities with the
same status.
Beyond these motivations for discriminating between individual and global 
fits, we add
our interest in investigating cosmic-ray information, which concerns only \st 
and not $\rho$.

\subsection{Cosmic ray information}

For $pp$ collisions, the total cross section may also be inferred from
cosmic-ray experiments and estimations exist in the high-energy interval
$\sqrt s = 6-40$ TeV. The procedure is model dependent and different 
analyses
lead to different results, as briefly reviewed in what follows.

The extraction of the proton-proton total cross section is based on the
 determination 
of the proton-air production
cross section from analysis of extensive air showers. Detailed reviews 
on the
subtleties involved may be found in Refs. \cite{gaisser,engel,gsy}.
Here we recall only the two main steps, stressing the model dependence
involved.

The first step concerns the determination of the proton-air production 
cross
section, namely, the ``inelastic cross section in which at least one 
new
hadron is produced in addition to nuclear fragments'' \cite{engel}. 
This is
obtained by the formulas

\begin{eqnarray}  
\sigma_{p\textrm{-}air}^{prod}\ (\textrm{mb}) = 
\frac{2.4 \times 10^4}{\lambda_{p\textrm{-}air}
\ (\textrm{g/cm}^2)},
\qquad \lambda_{p\textrm{-}air}\ (\textrm{g/cm}^2) = 
\frac{\lambda_{att}}{k}, \nonumber
\end{eqnarray} 
where $\lambda_{p\textrm{-}air}$ is the interaction length of protons in the
atmosphere, $\lambda_{att}$ is the shower attenuation length, and  
the inelasticity coefficient
$k$ is a measure of the dissipation of energy through the shower. 
$\lambda_{att}$ is an experimental quantity determined through the
$\chi_{max}$ attenuation method or the zenith angle attenuation technique
\cite{sokolsky}. On the other hand, the coefficient $k$ is model
dependent and obtained through
Monte Carlo simulation; roughly $k \approx 1.5 \rightarrow \ \approx 1$ 
($k > 1$) when going
from Feynman scaling models to strongly scaling violation models in the 
fragmentation
region \cite{engel}.

In a second step $\sigma_{tot}^{pp}$ is obtained from 
$\sigma_{p \textrm{-}air}^{prod}$
through the multiple diffraction formalism (MDF) by Glauber and Matthiae 
\cite{gm},
and taking into account the different processes and effects 
in the p-air total cross section:

\begin{eqnarray}  
\sigma_{p\textrm{-}air}^{tot} = \sigma_{p\textrm{-}air}^{prod} + 
\sigma_{p\textrm{-}air}^{el}  +
\sigma_{p\textrm{-}air}^{q\textrm{-}el} + \Delta \sigma, \nonumber
\end{eqnarray}
where $\sigma_{p\textrm{-}air}^{q-el}$ concerns the quasielastic excitation 
of the nucleus and
$\Delta \sigma$ is the Gribov screening correction due to multiple
scattering \cite{grib69}.

Two important inputs at this point are the nucleon distribution function 
and a
relation between the slope parameter $B$ and $\sigma_{tot}^{pp}$, necessary 
in the 
parametrization of the scattering amplitude:

\begin{eqnarray}  
F_{pp}(s,t) \propto \sigma_{tot}^{pp} \exp \left\{ \frac{B(s) t}{2}\right\}. 
\nonumber
\end{eqnarray} 
With the screening correction, the MDF allows the determination of all the 
above
cross sections and from the strong correlation between 
$\sigma_{p\textrm{-}air}^{prod} - \sigma_{tot}^{pp} - B$ the $pp$ total cross 
section may 
eventually be estimated (for more details, see \cite{engel,gsy}).

The results presently available from cosmic-ray experiments, in the above 
energy region,
are
characterized by discrepancies, mainly due to the model dependence of $k$ 
and
$B(s)$. Before reviewing these results, it is important to draw attention 
to
two facts. First, as is well known, discrepancies also characterize some 
\st data from
accelerator experiments at the
highest energies, for example, at $541-546$ GeV and mainly
at $1.8$ TeV. Second, when performing quantitative analyses, it is 
fundamental to
select as complete a set of information as possible, or at least those 
obtained
in similar circumstances or bases. In the case of cosmic-ray information, 
it is important
to stress that there is no experimental determination of $\sigma_{tot}^{pp}$,
 since all
the results are model dependent. Therefore they cannot be distinguished in 
terms of
an ``experimental'' status and {\it all} the information available must be 
considered.

Despite the model dependence involved,
we can classify the complete set of cosmic-ray information available 
according to the \textit{inputs or
procedures used} and, simultaneously, by the \textit{value of the total 
cross
section extracted}, as explained in what follows. 

On one hand the results usually quoted in the literature concern those
obtained by the Fly's Eye Collaboration in 1984 \cite{fly} and Akeno (AGASA)
Collaboration in 1993 \cite{akeno}.
In the first case, the authors used $k = 1.6$ (Feynman scaling model), a 
relation of proportionality
between $B$ and $\sigma_{tot}^{pp}$, namely, the geometrical scaling
model \cite{jorge}, and a Gaussian profile distribution for the nucleus. At 
$\sqrt s = 30$ TeV, 
from the value $\sigma_{p\textrm{-}air}^{prod} = 540 \pm 50$ mb, they extracted
$\sigma_{tot}^{pp} = 120 \pm 15$ mb
\cite{fly}. 
The Akeno Collaboration used $k = 1.5$ (Feynman scaling model) and the Durand 
and Pi approach
\cite{dp1} to extract the proton-proton cross sections in the region $\sqrt
s \sim 6-20$ TeV. We shall return to these results later on.

On the other hand, in 1987 Gaisser, Sukhatme, and Yodh (GSY) making use of the
Fly's Eye result for $\sigma_{p\textrm{-}air}^{prod}$, as an estimate of the 
\textit{lowest}
allowed values for $\sigma_{tot}^{pp}$, and the Chou-Yang prescription
for $B(s)$ \cite{cy83} obtained $\sigma_{pp}^{tot} = 175_{-27}^{+40}$ mb 
at
$\sqrt s = 40$ TeV \cite{gsy}.
In 1993, Nikolaev claimed that the Akeno results should be corrected
in order to take into account the differences between absorption and
inelastic cross sections, leading to an increase of the results by
$\approx 30$ mb \cite{niko}. The analysis by Nikolaev was also motivated 
by
previous results from a QCD model of the Pomeron \cite{knp}.

At this point some fundamental facts concerning all these results must be
stressed. First, as reviewed above, the Fly's Eye and Akeno results for
$\sigma_{tot}^{pp}$ are as model dependent as those by Nikolaev and GSY
and therefore cannot be considered as experimental results; what can
and must be discussed is the class of model used in each case.
In this sense, we first note that since the Fly's Eye Collaboration
used the geometrical scaling hypothesis, which is violated even at the
collider energy, their result is probably wrong. Second, in 1990, Durand 
and Pi
asserted in Ref. \cite{dp2} that their results published in 1988 \cite{dp1},
and used by the Akeno Collaboration \cite{akeno}, should be disregarded due 
to a wrong 
approximation concerning fluctuations. 
The new results presented in \cite{dp2} (Fig. 11 in that paper) introduced 
significant changes 
in the scenario.
For example, for $\sigma_{p-air} \approx 550$ mb, the Akeno result with the method
by Durand and Pi \cite{dp1} was $\sigma_{tot}^{pp} = 124$ mb and from Fig. 
11 in Ref. 
\cite{dp2} the value extracted is $\sigma_{tot}^{pp} = 137$ mb, that is, an 
increase of 
$\approx 11 \% $; to the Fly's Eye published point \cite{fly}, the corresponding 
value is
$\sigma_{tot}^{pp} = 135$ mb \cite{dp2}, that is, an increase of $\approx 13 \% 
$.
However, to our knowledge no other publication by Durand and Pi on the subject 
appeared 
in the literature.
Finally, it should be stressed that the Chou-Yang prescription,
used by Gaisser, Sukhatme, and Yodh, is somewhat model independent and, 
also to our knowledge, the results by these authors together with those by
Nikolaev have never been criticized, despite the intrinsic model dependence 
involved.
In conclusion, we understand that all these facts, not usually discussed in 
the
literature, suggest an increase of the total cross section faster than 
indicated
by the Akeno and Fly's Eye results and that this indication has a reasonable
basis.

It should also be recalled that, more recently, Block, Halzen, and Stanev 
(BHS)
obtained estimations for $\sigma_{tot}^{pp}$ from $\sigma_{p\textrm{-}air}$ \cite{bhs},
through a QCD-inspired model \cite{qcdim} and the MDF.
In the analysis, the inelasticity coefficient $k$ is considered as a free
parameter, determined through a global fit, including 
$pp$ and $\bar{p}p$ accelerator data and the published results on the
$p\textrm{-}air$ cross section \cite{bhs}. Specifically, $k$ is determined in a way that 
their
predictions for $\sigma_{tot}^{pp} = \sigma_{tot}^{\bar{p}p}$ (asymptotically) 
matches 
$\sigma_{p\textrm{-}air}$ through the MDF. 
The resulting value $k \approx 1.35$ seems to
be in accord with those obtained by combinations of different simulation
programs, namely, $k \approx 1.15-1.30$ \cite{pryke}. The extracted
$\sigma_{tot}^{pp}(s)$ at the cosmic-ray energies shows agreement with the 
Akeno
results and is about $17$ mb below the Fly's Eye value at $30$ TeV. 

The numbers associated with all the above cosmic-ray estimations of 
$\sigma_{tot}^{pp}$ 
are displayed in Table 1.
Despite the  importance and originality of the approach by BHS, 
we shall not include it in our analysis, because they are in agreement with 
the
Akeno results (Table 1) that will be used.
However we will treat this case in  further work.
With the exception of the BHS results,
all the other cosmic-ray estimates discussed above together with the 
accelerator 
results are displayed in Fig. 1.

\begin{table*}
\caption{Estimations for  $\sigma_{tot}^{pp}(s)$ (in mb) from cosmic-ray data
\cite{pdg}.}
\begin{ruledtabular}
\begin{tabular}{ccccccc}
\multicolumn{1}{c}{$E_{lab}$} & \multicolumn{1}{c}{$\sqrt{s}$} & 
\multicolumn{1}{c}{Akeno (AGASA)} & \multicolumn{1}{c}{Fly's Eye} & 
\multicolumn{1}{c} {Nikolaev}
& \multicolumn{1}{c} {GSY}& \multicolumn{1}{c} {BHS} \\
($10^{16}$ eV) & (TeV) & \cite{akeno} &  \cite{fly} &  
\cite{niko}& \cite{gsy}& \cite{bpc} \\
\hline
2.02 & 6.2 & 93$\pm$14 & - & 120$\pm$15 & - & 91$\pm$ 15\\
3.52 & 8.1 & 101$\pm$16 & - & 130$\pm$18 & - & 100$\pm$ 18\\
6.11 & 10.6 & 117$\pm$18 & - & 154$\pm$17 & - & 118$\pm$ 17\\
10.63 & 14.0 & 104$\pm$26 & - & 135$\pm$29 & - & 103$\pm$ 29\\
18.47 & 18.4 & 100$\pm$27 & - & 129$\pm$30 & - & 99$\pm$ 30\\
32.09 & 24.3 & 124$\pm$34 & - & 162$\pm$38 & - & 124$\pm$ 37\\
47.96 & 30.0 & - & 120$\pm$15 & - & - & 103$\pm$ 22\\
85.26 & 40.0 & - & - & - & 175$\pm$34 & -\\
\end{tabular}
\end{ruledtabular}
\end{table*}

\begin{figure}[ht]
\includegraphics[width=12cm,height=11cm]{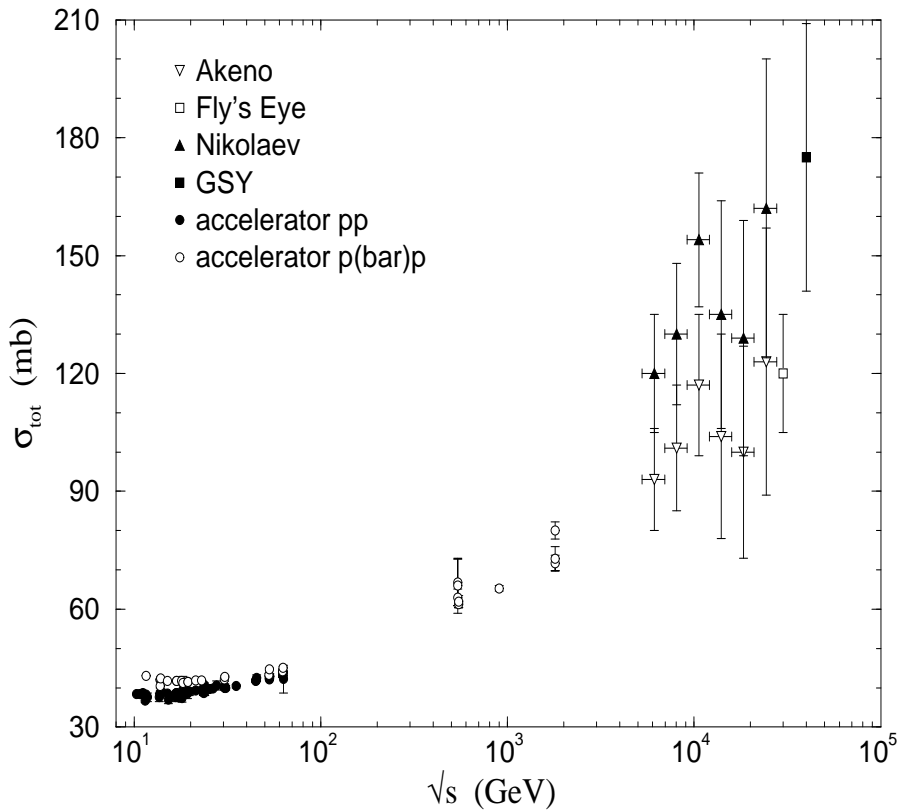}
\caption{Total cross sections ($pp$ and $\overline{p}p$ above
$\sqrt s = 10$\ GeV): accelerator data and cosmic-ray information 
available
(Table I).}
\label{alm01f1}
\end{figure}

\subsection{Ensembles}

From Fig. 1 we see that, despite the large error bars in
the cosmic-ray region, we can
identify two distinct set of estimations: one represented by the results 
of
the Fly's Eye Collaboration together with those by the Akeno
Collaboration; the other by the results of Gaisser, Sukhatme, and Yodh
with those by Nikolaev, which follow the higher estimates by Durand 
and Pi
\cite{dp2}. Taken separately these two sets suggest different
scenarios for the increase of the total cross section, as already 
claimed 
before \cite{bjp01,alm,mm}.

Based on these considerations, we shall investigate the behavior of the
total cross section by taking into account  the discrepancies that
characterize the cosmic ray information and, 
to this end, we shall consider \textit{two ensembles of data and
experimental information} with the following notation:

\vspace{0.3cm}

$\bullet$ \textit{Ensemble I}:
$\bar{p}p$ accelerator data and $pp$ accelerator data + Akeno +
 Fly's Eye;

\vspace{0.3cm}

$\bullet$ \textit{Ensemble II}:
$\bar{p}p$ accelerator data and $pp$ accelerator data + Nikolaev + 
GSY.

\vspace{0.3cm}

To some extent, ensemble I represents a kind of high-energy standard 
picture
and ensemble II a nonstandard one. However, the consistence among  the  
results by Nikolaev and GSY must be stressed , in addition to their agreement 
with the last results by Durand and Pi (Fig. 11 in Ref. \cite{dp2}). 
We add also that, from
the discussion in Sec. II B, both ensembles seem equally probable.

\section{Analytical Approach}

In this section we first review the essential formulas in the derivative 
dispersion
approach and recall some fundamental theorems and high-energy bounds. 
The parametrizations that characterize the analytical models to be used
in the next section are also presented.

\subsection{Analyticity relations}

For $pp$ and $\overline{p}p$ scattering, analyticity and crossing symmetry
allow us to connect $\sigma_{tot}(s)$ and $\rho(s)$ through two compact and
symmetric formulas:  
\begin{eqnarray}
\rho ^{pp} \sigma_{tot}^{pp} (s) =
E( \sigma_{+} ) + O( \sigma_{-} ),
\end{eqnarray}

\begin{eqnarray}
\rho ^{\bar{p}p} \sigma_{tot}^{\bar{p}p} (s) =
E( \sigma_{+} ) - O( \sigma_{-} ),
\end{eqnarray}
where

\begin{eqnarray}  
\sigma_{\pm}(s) = \frac{\sigma_{tot}^{pp} \pm 
\sigma_{tot}^{\bar{p}p}}{2},
\end{eqnarray} 
and $E( \sigma_{+} )$,$O( \sigma_{-} )$ are analytic transforms 
relating the
real and imaginary parts of crossing even and odd amplitudes, 
respectively.
These analyticity relations are usually expressed in an \textit{integral} 
form and
in the case of the forward direction the standard \textit{once subtracted} 
integral dispersion
relations may be expressed by \cite{idr,bc}
 
\begin{eqnarray} 
E_{int}(\sigma_{+}) \equiv
\frac{K}{s} + \frac{2s}{\pi}\int_{s_o}^{\infty} 
\mathrm{d}s' \left[\frac{1}{s'^2-s^2}\right]  
\sigma_{+}(s') = \frac{\textrm{Re}\ F_{+}(s)}{s},
\end{eqnarray}

\begin{eqnarray} 
O_{int}(\sigma_{-}) \equiv
  \frac{2}{\pi}\int_{s_o}^{\infty} 
\mathrm{d}s' \left[\frac{s'}{s'^2-s^2}\right]  
\sigma_{-}(s') = \frac{\textrm{Re}\ F_{-}(s)}{s},
\end{eqnarray}
where $K$ is the subtraction constant. Since we shall be interested in the 
high-energy
region, we used in the above formulas the c.m. energy, instead of the
 laboratory 
energy and momentum \cite{idr,bc}.

   At sufficiently high energies integral analyticity
relations may be replaced by derivative forms, usually called 
\textit{analyticity relations},
 which are more
useful for practical calculations. In these formulas, differentiation with
respect to the logarithm of the energy occurs in the argument of a
trigonometric operator expressed by its series \cite{ddr,bks,kn}.
Here we shall use the standard form deduced by Bronzan, Kane, and Sukhatme 
\cite{bks}
(see also \cite{bjp01,mmp}), obtained from the integral form
in the high-energy limit:

\begin{eqnarray} 
E_{der}(\sigma_{+}) \equiv
\frac{K}{s} + 
\tan\left[\frac{\pi}{2}\frac{d}{d\ln s} \right] 
\sigma_{+}(s) =  \frac{\textrm{Re}\ F_{+}(s)}{s}, \nonumber
\end{eqnarray}

\begin{eqnarray} 
O_{der}(\sigma_{-}) \equiv
\tan\left[\frac{\pi}{2}\left( 1 + \frac{d}{d\ln s}\right) \right] 
\sigma_{-}(s) =  \frac{\textrm{Re}\ F_{-}(s)}{s}. \nonumber
\end{eqnarray} 

Operationally these transforms may be evaluated through the expansions

\begin{eqnarray} 
E_{der}(\sigma_{+}) - \frac{K}{s} =
\left[ \frac{\pi}{2} \frac{d}{d\ln s} + 
\frac{1}{3} \left(\frac{\pi}{2}\frac{d}{d \ln s}\right)^3 +
\frac{2}{5} \left(\frac{\pi}{2}\frac{d}{d \ln s}\right)^5 + \
.\ .\ .\ \right] \sigma_{+}(s),
\end{eqnarray}

\begin{eqnarray} 
O_{der}(\sigma_{-})  &=& 
- \int \left\{ \frac{d}{d\ln s} \left[\cot \left( \frac{\pi}{2} 
\frac{d}{d\ln s} \right)\right] \sigma_{-}(s) \right\} d\ln s \nonumber \\
&=&
- \frac{2}{\pi}\int \left\{ \left[ 1 - \frac{1}{3} \left(\frac{\pi}{2}\frac{d}{d \ln
s}\right)^2 -  \frac{1}{45} \left(\frac{\pi}{2}\frac{d}{d \ln s}\right)^4
 -\ .\ .\ .\ \right] \sigma_{-}(s) \right\} d \ln s.
\end{eqnarray}

This completes the analytical approach: with an input parametrization for 
$\sigma_{tot}(s)$,
Eqs. (1)-(4) allow, in principle, the determination of $\rho(s)$, by
 means
of either the integral forms Eqs. (5)-(6) or the derivative (analyticity) ones 
Eqs. (7)-(8).

\subsection{Theorems at high energies}

For future reference, we briefly recall here two rigorous high-energy results
for the asymptotic behavior of the cross sections.
The Froissart-Martin bound states that as $s \rightarrow \infty$
\cite{froi,mart}

\begin{eqnarray} 
\sigma_{tot} \leq C \ln^2 s, \nonumber
\end{eqnarray} 
and, according to the generalized form of the Pomeranchuk theorem 
\cite{eden,grutru},
if the Froissart-Martin bound is reached the difference between 
$\overline{p}p$ and $pp$ total cross sections goes as

\begin{eqnarray} 
\Delta \sigma_{tot} =\sigma_{tot}^{p\overline{p}} -  \sigma_{tot}^{pp}
\leq c \frac{\sigma_{tot}^{\bar{p}p} + \sigma_{tot}^{pp}}{\ln s}, 
\nonumber
\end{eqnarray} 
which means that the difference may increase as $\ln s$.
From Eq. (4), this difference is given by the crossing odd component

\begin{eqnarray} 
\Delta \sigma_{tot}(s) = -2\sigma_{-}(s), \nonumber
\end{eqnarray} 
and therefore, if the Froissart-Martin bound is saturated, a rigorous
 result is
that, asymptotically, the difference  $\Delta
\sigma_{tot}(s)$ is controlled by the odd component and the maximum
contribution is given by $\sigma_{-}^{max}(s) = \ln s$. This corresponds 
to one
of the variants of the Odderon
picture \cite{odd,kn} and the increase as $\ln s$ is the maximum Odderon
hypothesis. If the odd contribution in the imaginary part of the amplitude
is not present at the highest energies, then $\Delta \sigma_{tot} = 0$. The
possible effects in the real part will be discussed in what follows.

\subsection{Analytic models}

In the formulas that follows we denote $s/s'$ with $s' = 1$ GeV$^2$ by
$s$. We shall consider two different parametrizations for the total cross
section, the main difference being the asymptotic limits, which allow  the
dominance of an even or odd amplitude.

\subsubsection{ Donnachie-Landshoff}

The Donnachie-Landshoff (DL) parametrizations for the total cross sections
are expressed by \cite{dl}

\begin{eqnarray}
\sigma_{tot}^{pp} (s) = X s^{\epsilon} + Y s^{- \eta},  
\qquad
\sigma_{tot}^{\bar{p}p} (s) = X s^{\epsilon} + Z s^{- \eta}, 
\end{eqnarray} 
where, originally, $X = 21.7$ mb, $Y = 56.08$ mb, $Z = 98.39$ mb, 
$\epsilon = 0.0808$, and $\eta = 0.4525$. We observe that this model 
predicts that the difference 
between the two cross sections is given by

\begin{eqnarray}
\Delta \sigma = \sigma^{\overline{p}p}_{tot}(s) - \sigma^{pp}_{tot}(s) =
 (Z - Y) s^{-\eta} \rightarrow 0 \quad {\rm (asymptotically)}.\nonumber
\end{eqnarray}

Through the formalism described in Sec. III A, substitution of the
 parametrizations (9)
into Eq. (4) and then in Eqs. (5)-(6) with $s_0 = 0$ or 
Eqs. (7)-(8) gives

\begin{eqnarray}
E(\sigma_{+}) = \frac{K}{s} +
\left[X\tan\left(\frac{\pi\epsilon}{2}\right)\right]s^{\epsilon} -
\left[\frac{(Y+Z)}{2}\tan\left(\frac{\pi\eta}{2}\right)\right]s^{\eta},
\nonumber
\end{eqnarray}

\begin{eqnarray}
O(\sigma_{-}) =
\left[\frac{(Y-Z)}{2}\cot\left(\frac{\pi\eta}{2}\right)\right]s^{\eta}, \nonumber
\end{eqnarray}
and from Eqs. (2)-(3) we obtain the analytical expressions for $\rho(s)$:

\begin{eqnarray}
\rho ^{pp}(s) = \frac{1}{\sigma_{tot}^{pp}(s)}
\left\{ \frac{K}{s} +
\left[X\tan\left(\frac{\pi\epsilon}{2}\right)\right]s^{\epsilon} +
\left[\frac{(Y-Z)}{2}\cot\left(\frac{\pi\eta}{2}\right) -
\frac{(Y+Z)}{2}\tan\left(\frac{\pi\eta}{2}\right)\right]s^{- \eta}\right\}, 
\end{eqnarray}

\begin{eqnarray}
\rho ^{\overline{p}p}(s) = \frac{1}{\sigma_{tot}^{\overline{p}p}(s)}
\left\{ \frac{K}{s} +
\left[X\tan\left(\frac{\pi\epsilon}{2}\right)\right]s^{\epsilon} +
\left[\frac{(Z-Y)}{2}\cot\left(\frac{\pi\eta}{2}\right) -
\frac{(Y+Z)}{2}\tan\left(\frac{\pi\eta}{2}\right)\right]s^{- \eta}\right\}.
\end{eqnarray}

Since in this model $\Delta \sigma \rightarrow 0$ asymptotically, for
$\sigma_{tot}^{\bar{p}p} = \sigma_{tot}^{pp} \equiv \sigma_{tot}$ we have

\begin{eqnarray}
\Delta \rho = \rho^{\bar{p}p} -  \rho^{pp} \sim \frac{1}{\sigma_{tot}(s)}
(Z-Y)\cot\left(\frac{\pi\eta}{2}\right)s^{-\eta} \quad \rightarrow \quad 0
\qquad {\rm as} \qquad s \quad \rightarrow \quad \infty. \nonumber
\end{eqnarray}

\subsubsection{Kang-Nicolescu}

The parametrizations for the total cross sections used by Kang and 
Nicolescu (KN) under the 
hypothesis of the Odderon \cite{odd} are expressed by \cite{kn}
\begin{eqnarray}
\sigma^{pp}_{tot}(s)= A_{1} + B_{1} \ln s  +
k \ln^2 s,
\qquad
\sigma^{\overline{p}p}_{tot}(s)= A_{2} + B_{2} \ln s 
+ k \ln^2 s + 2Rs^{-1/2}.
\label{ppbar}
\end{eqnarray}

Differently from the previous case, this model predicts that the difference
 between the two 
cross sections is given by

\begin{eqnarray}
\Delta \sigma = \sigma^{\overline{p}p}_{tot}(s) &-& \sigma^{pp}_{tot}(s) =
(A_2 - A_1) + (B_2 - B_1)\ln s + 2Rs^{-1/2} \\
&\rightarrow& \Delta A + 
\Delta B \ln s \quad {\rm(asymptotically)}, \nonumber
\end{eqnarray}
so that, if $\Delta A
\not= 0$ and/or  $\Delta B \not= 0$, the total cross section difference may 
increase and 
$\sigma_{tot}^{pp}$ may even become greater than $\sigma_{tot}^{\bar{p}p}$, 
depending on the
values and signs of $\Delta A$ and $\Delta B$, which is formally in 
agreement with
the theorems of Sec. III B.

With a similar procedure as in the previous model the use of the 
\textit{analyticity relations} (7)-(8) leads to

\begin{eqnarray}
E(\sigma_{+}) = \frac{K}{s} +
\frac{\pi}{2} \left(\frac{B_1 + B_2}{2}\right) + \pi k \ln s -R
s^{-1/2},\nonumber \end{eqnarray}

\begin{eqnarray}
O(\sigma_{-}) =  
\left( \frac{A_2 - A_1}{\pi} \right) \ln s + \left(\frac{B_2 - B_1}{2\pi}
\right) \ln ^2 s  - R s^{-1/2}, \nonumber
\end{eqnarray} 

and from Eqs. (2)-(3)

\begin{eqnarray}
\rho ^{pp} = \frac{1}{\sigma_{tot}^{pp}}
\left\{ \frac{K}{s} + \frac{\pi}{2}\left(\frac{B_1 + B_2}{2}\right) + 
\left(\pi
k + \frac{A_2 - A_1}{\pi}\right) \ln s + \left(\frac{B_2 - B_1}{2\pi} 
\right)
\ln ^2 s  - 2Rs^{-1/2} \right\},
\end{eqnarray}

\begin{eqnarray}
\rho ^{\overline{p}p} = \frac{1}{\sigma_{tot}^{\overline{p}p}}
\left\{ \frac{K}{s} + \frac{\pi}{2}\left(\frac{B_1 +
B_2}{2}\right) + \left(\pi k - \frac{A_2 - A_1}{\pi}\right)\ln s -
\left(\frac{B_2 - B_1}{2\pi} \right) \ln ^2 s  \right\}. 
\end{eqnarray}

In this case, if $\Delta A$ and $\Delta B$ are sufficiently small, so that
we may replace $\sigma_{tot}^{\bar{p}p} \approx \sigma_{tot}^{pp} \equiv 
\sigma_{tot}(s)$, then, 
asymptotically,

\begin{eqnarray}
\Delta \rho = \rho ^{\overline{p}p}  - \rho ^{pp} \sim
- \frac{1}{\pi \sigma_{tot}(s)} \left\{ \Delta A \ln s + \Delta B \ln^2 s 
\right\}.
\end{eqnarray}
This means that, depending on the fit results, there may be a change of 
sign in
$\Delta \rho$, with $\rho ^{pp}$ becoming greater than 
$\rho ^{\overline{p}p}$ at some
finite energy.

All these possibilities for a change of sign in the differences 
between $\bar{p}p$
and $pp$ total cross sections and/or the $\rho$ parameters 
are based on the concept of the Odderon \cite{odd}; the case of a 
change of sign
in $\Delta \sigma_{tot}(s)$ was early discussed by Bernard, Gauron, and 
Nicolescu \cite{bgn}
and that in $\Delta  \rho$ by Gauron, Nicolescu, and Leader \cite{gnl}. 
They are associated
with the condition that the maximal odderon dominates the imaginary or 
real part
of the amplitude, respectively.

\section{Fitting of the data and results}

In order to investigate all the points raised in Sec. I and their 
possible interconnections, 
we perform 16 different fits through the program CERN-MINUIT. In these 
fits we use both
ensembles I and II defined in Sec. II and both the DL and KN models 
described in
Sec. III. For each of these four possibilities we perform global and 
individual
fits to \st and \ro and, in each case, we either consider the 
subtraction
constant $K$ as a free fit parameter or assume $K = 0$. 

In the individual approach we first fit only the total cross section 
data and then
extract the corresponding $\rho(s)$ in the case of $K=0$, or, with 
the results
for $\sigma_{tot}(s)$, we fit only the $\rho$ data with $K$ as a free 
fit parameter.

The  numerical results of the fits and statistical information are 
all displayed
in Tables II and III for the DL  and KN models, respectively. The 
corresponding
curves together with the experimental data are shown in Figs. 2, 3, 
and 4 for the
DL model and Figs. 5, 6, and 7 for the KN model.

\begin{table*}
\caption{Results with the Donnachie-Landshoff model.} 
\label{dl}
\begin{ruledtabular}
\begin{tabular}{ccccccccc} 
Fit: & \multicolumn{4}{c}{Individual } & \multicolumn{4}{c}{Global} \\
Quantity: & \multicolumn{2}{c}{\st} & \multicolumn{2}{c}{\ro} 
 & \multicolumn{4}{c}{\st and \ro}\\ 
Ensemble:  &  I &  II &  I &  II  &  I &  II &  I &  II\\ 
\hline
No. of points  &   $102$    &   $102$ & $63$ &$63$ & $165$ & $165$ & 
165& 165\\  
$\chi^2/\textrm{DOF}$   &  $0.76$ &  $0.96$ & 1.55&1.84 & $1.09$ & 
$1.24$ & 0.84 & 0.98\\  
 $X$ (mb) & $ 20.0 \pm 0.7$  & $19.3 \pm 0.7$ & - & - & $21.6 \pm 0.4$ & 
$ 21.2 \pm
0.4$ & 21.4 $\pm$ 0.4& 21.1 $\pm$ 0.4 \\   
$Y$ (mb) & $ 48 \pm 5$  & $46 \pm 5$ & - & - & $51 \pm 3 $ & $51 \pm 3$ &
67 $\pm$6 &67 $\pm$ 5 \\  
$Z$ (mb) & $74 \pm 10$ & $70 \pm 9$ & - & - & $85 \pm 5$ & $84 \pm 5$ &
114 $\pm$ 11 & 112 $\pm$ 11\\ 
$\eta$  & $0.37 \pm 0.03$ &  $0.35 \pm 0.03$ & - & - & $0.43 \pm 0.02$ & 
$0.42
\pm 0.02$ &0.48 $\pm$ 0.02 & 0.47 $\pm$ 0.02\\  
$\epsilon$ & $0.088 \pm 0.003$  &  $0.091 \pm 0.003$ & - & - & 
$0.081 \pm 0.002$ &
$0.083 \pm 0.002$ &0.083 $\pm$ 0.002 & 0.084 $\pm$ 0.002\\  
\hline
$K$ &  -  &  - & 235 $\pm$ 32 & 245 $\pm$ 32 & 0 & 0 &306 $\pm$ 54 & 
307 $\pm$ 52\\ 
\hline
Figure: & \multicolumn{2}{c} {2(a) and 2(b)} & \multicolumn{2}{c}{2(a) 
and 2(c)} 
 & \multicolumn{2}{c} {3} & \multicolumn{2}{c}{4}\\ 
\end{tabular}
\end{ruledtabular}
\end{table*}

\begin{figure}
\psfig{file=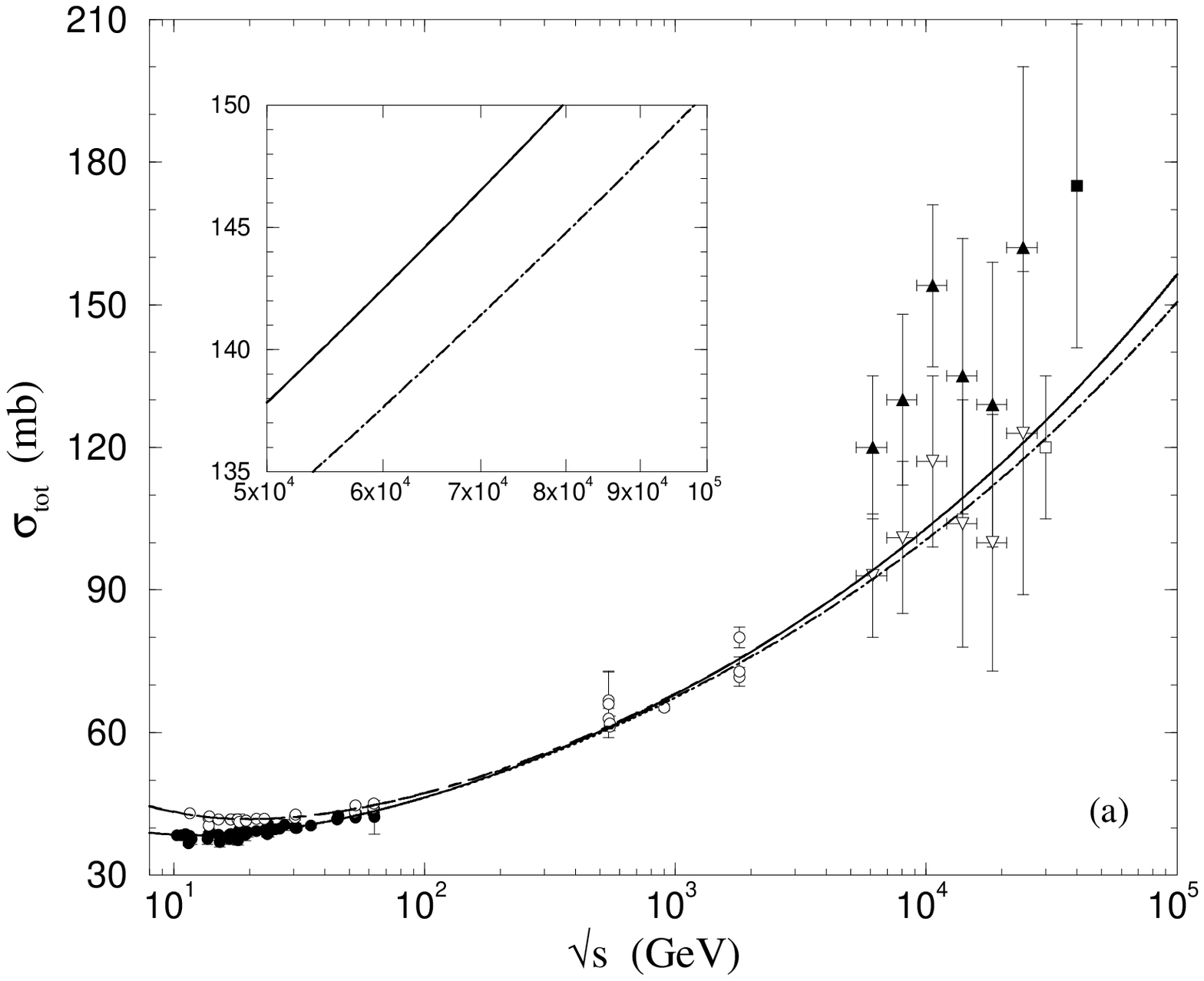,width=5.9cm,height=9cm}
\psfig{file=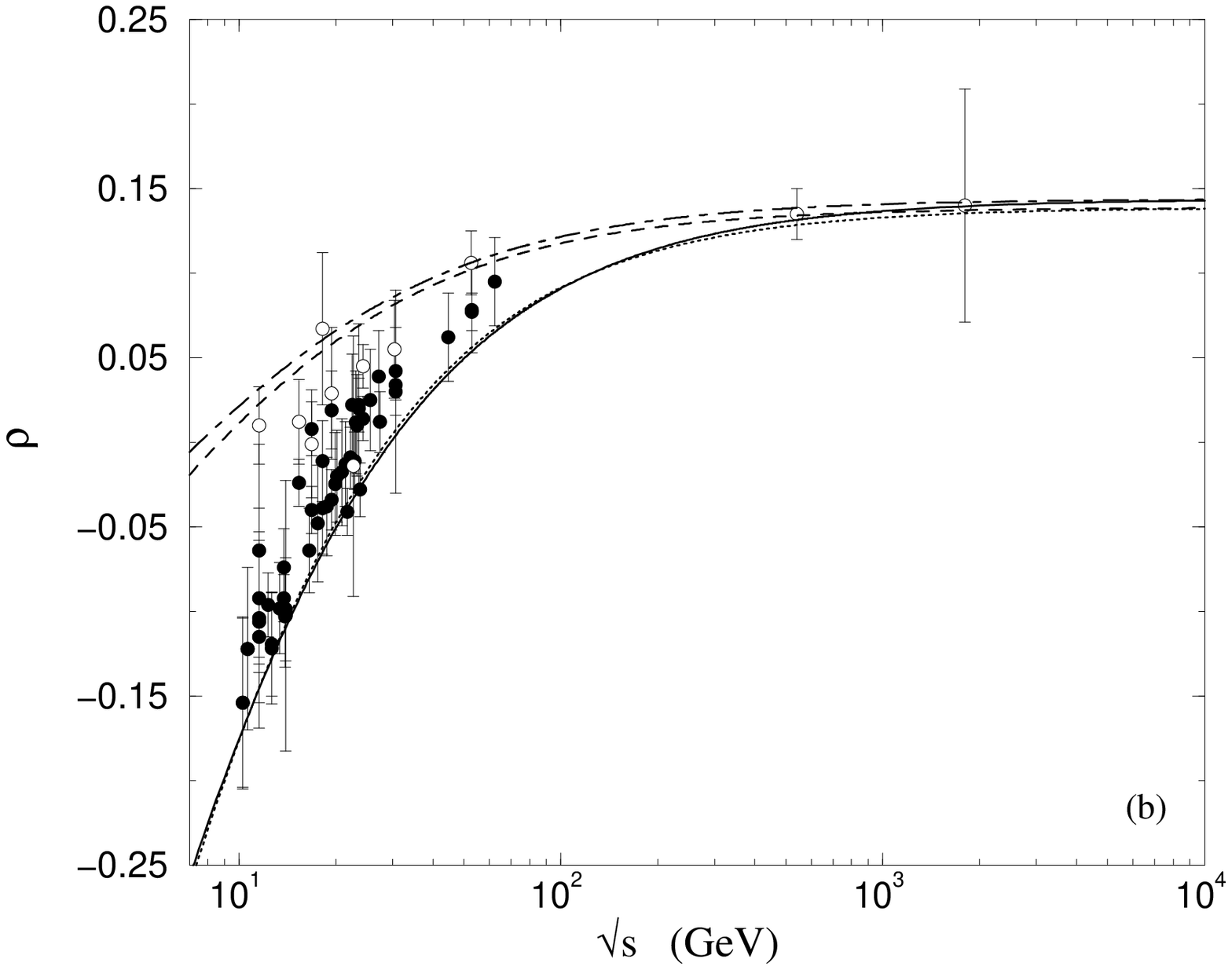,width=5.9cm,height=9cm}
\psfig{file=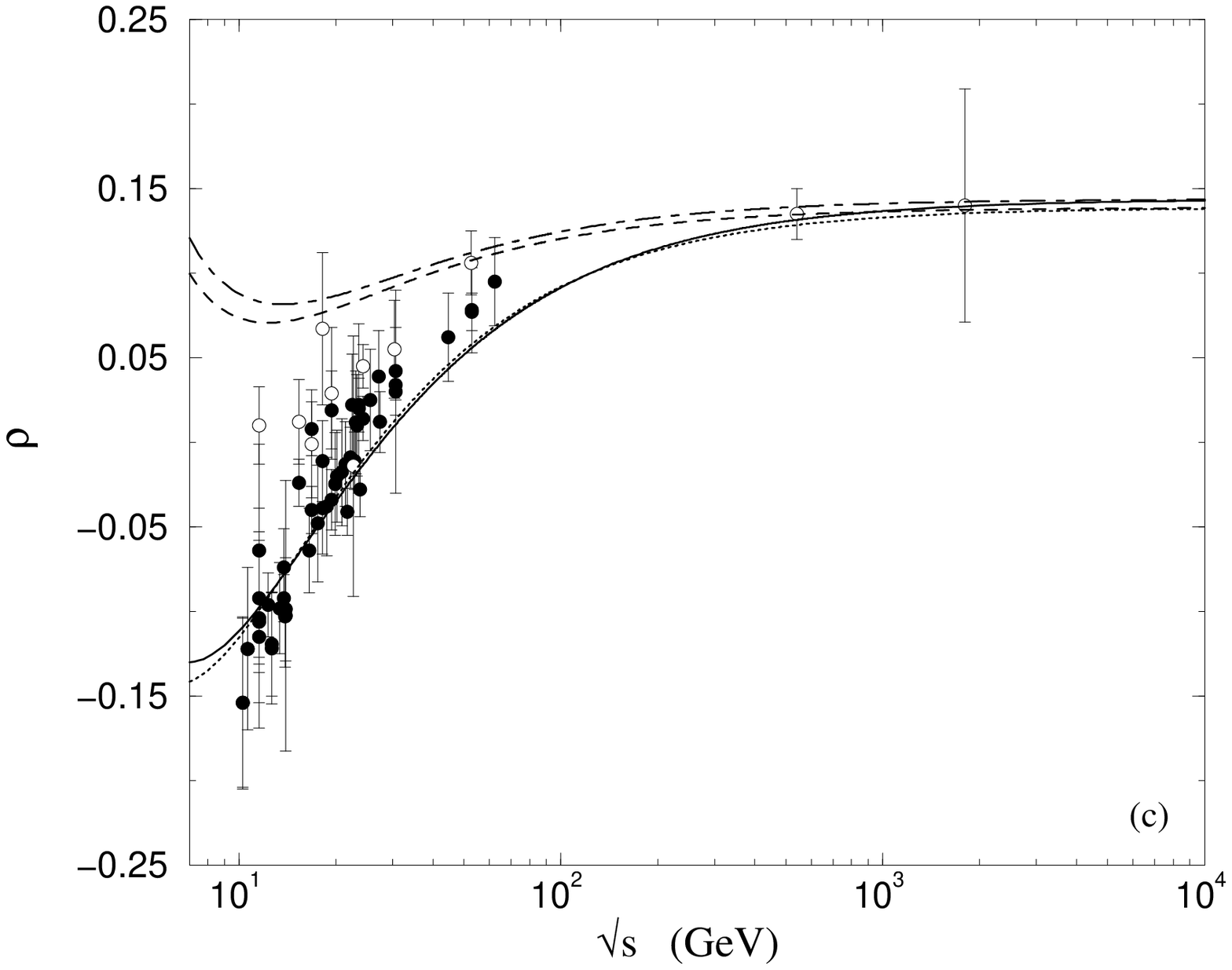,width=5.9cm,height=9cm}
\caption{Fits to $pp$ and $\overline{p}p$ total cross section data from
ensembles I (dotted curves for $pp$ and dashed for $\bar{p}p$) and II 
(solid curves for $pp$ and dot-dashed for $\bar{p}p$), 
through the DL parametrization (a) and the
corresponding predictions for $\rho(s)$ with $K=0$ (b) and $K$ as
 free fit parameter
(c)---columns 2--5 in Table II.}
%\label{rates}
\end{figure}

\begin{figure}
\psfig{file=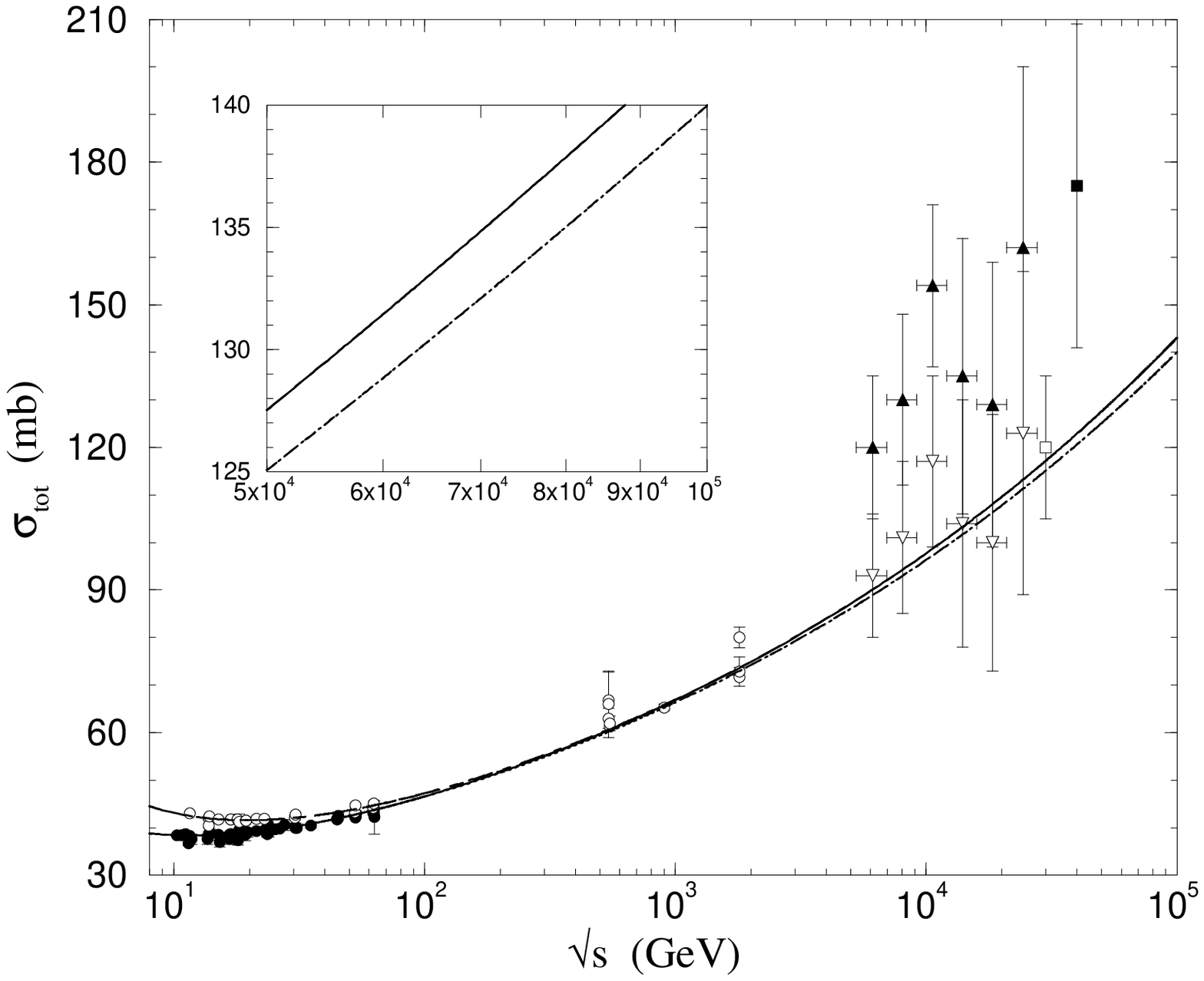,width=8.5cm,height=10cm}
\psfig{file=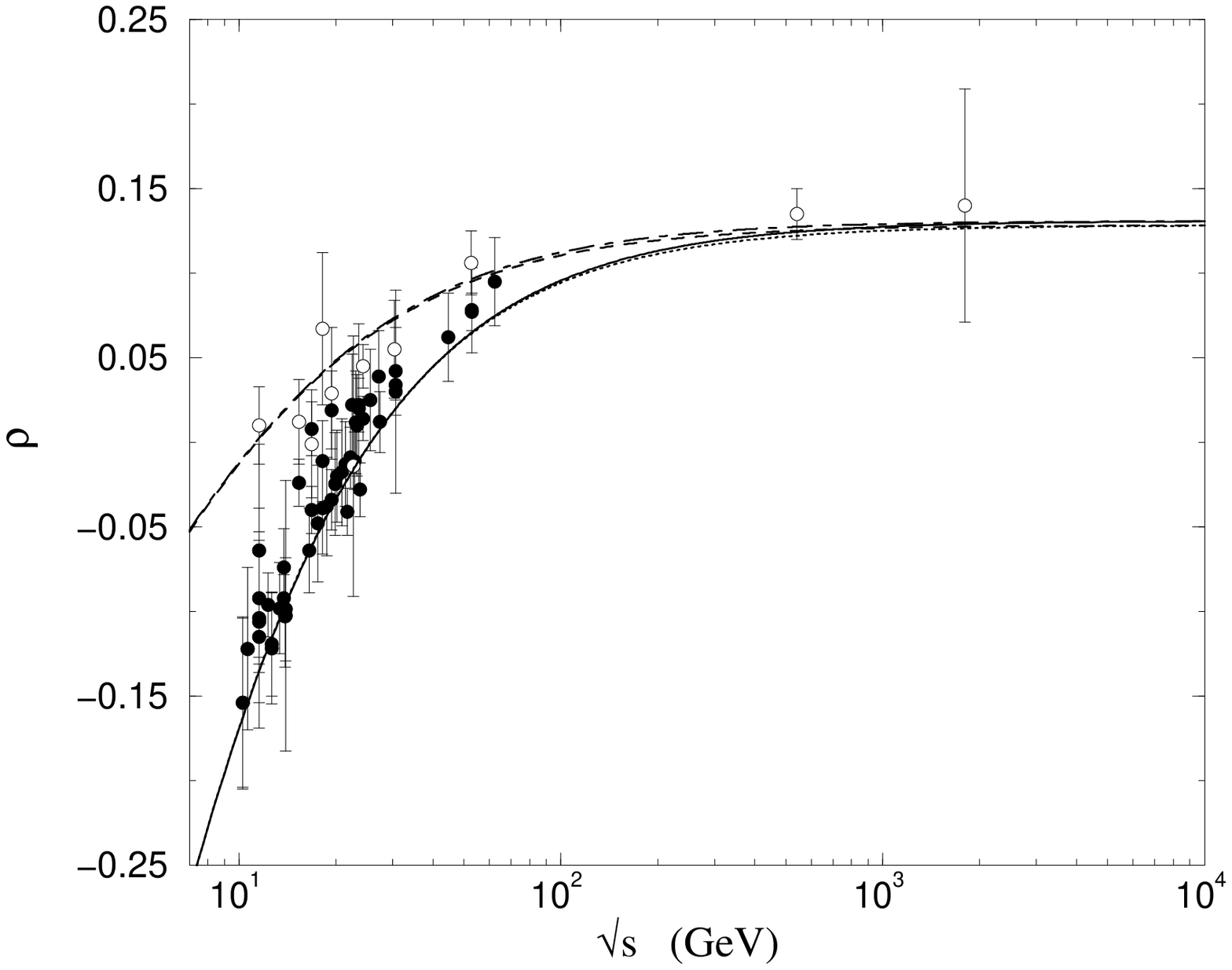,width=8.5cm,height=10cm}
\caption{\noindent
Simultaneous fits to $\sigma_{tot}(s)$ and $\rho(s)$ through the DL
parametrization with $K=0$ and ensembles I (dotted curves for $pp$ and 
dashed for $\bar{p}p$) and II 
(solid curves for $pp$ and dot-dashed for $\bar{p}p$)---columns 6 and 7
 in Table II.}
%\label{rates}
\end{figure}

\begin{figure}
\psfig{file=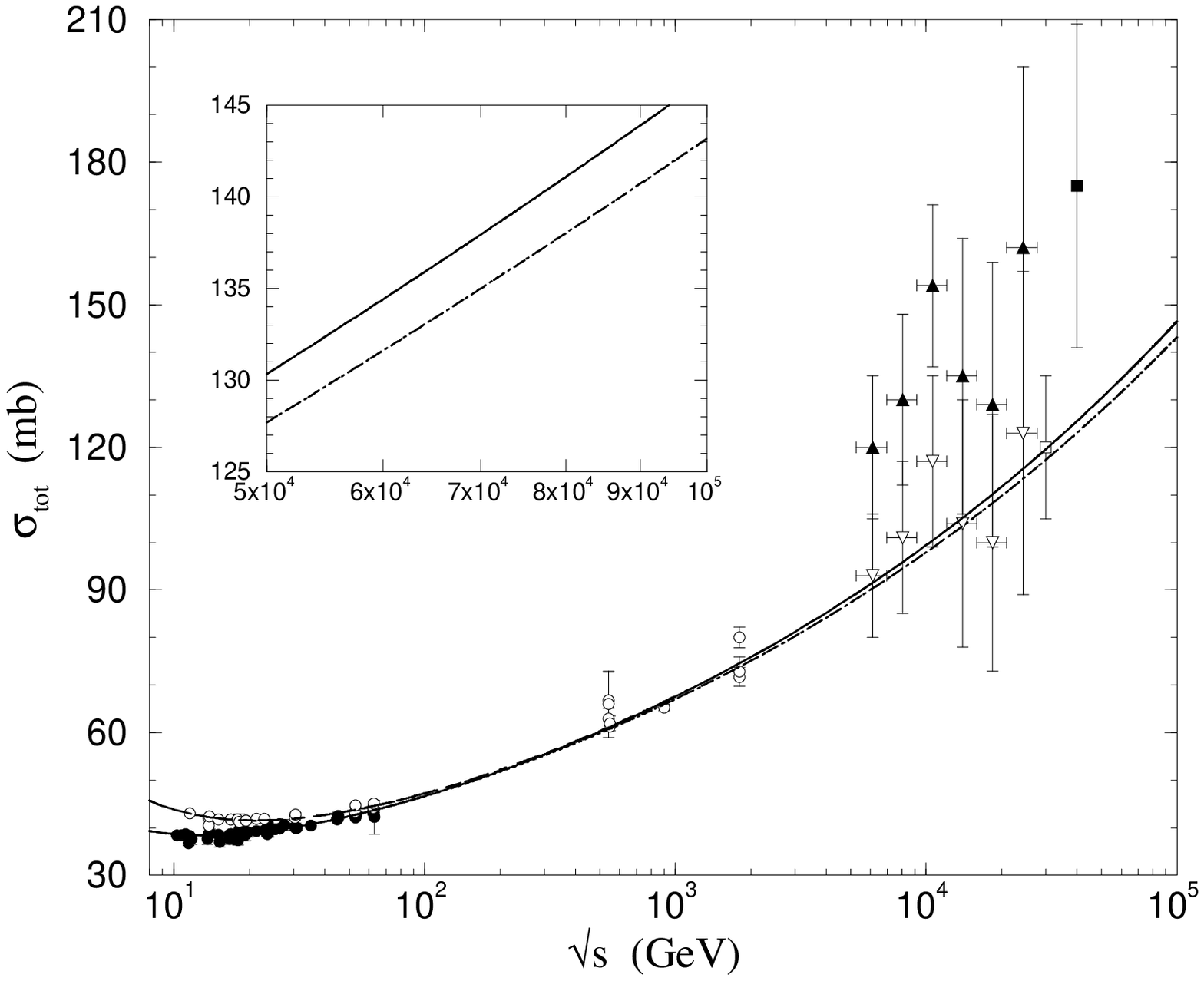,width=8.5cm,height=10cm}
\psfig{file=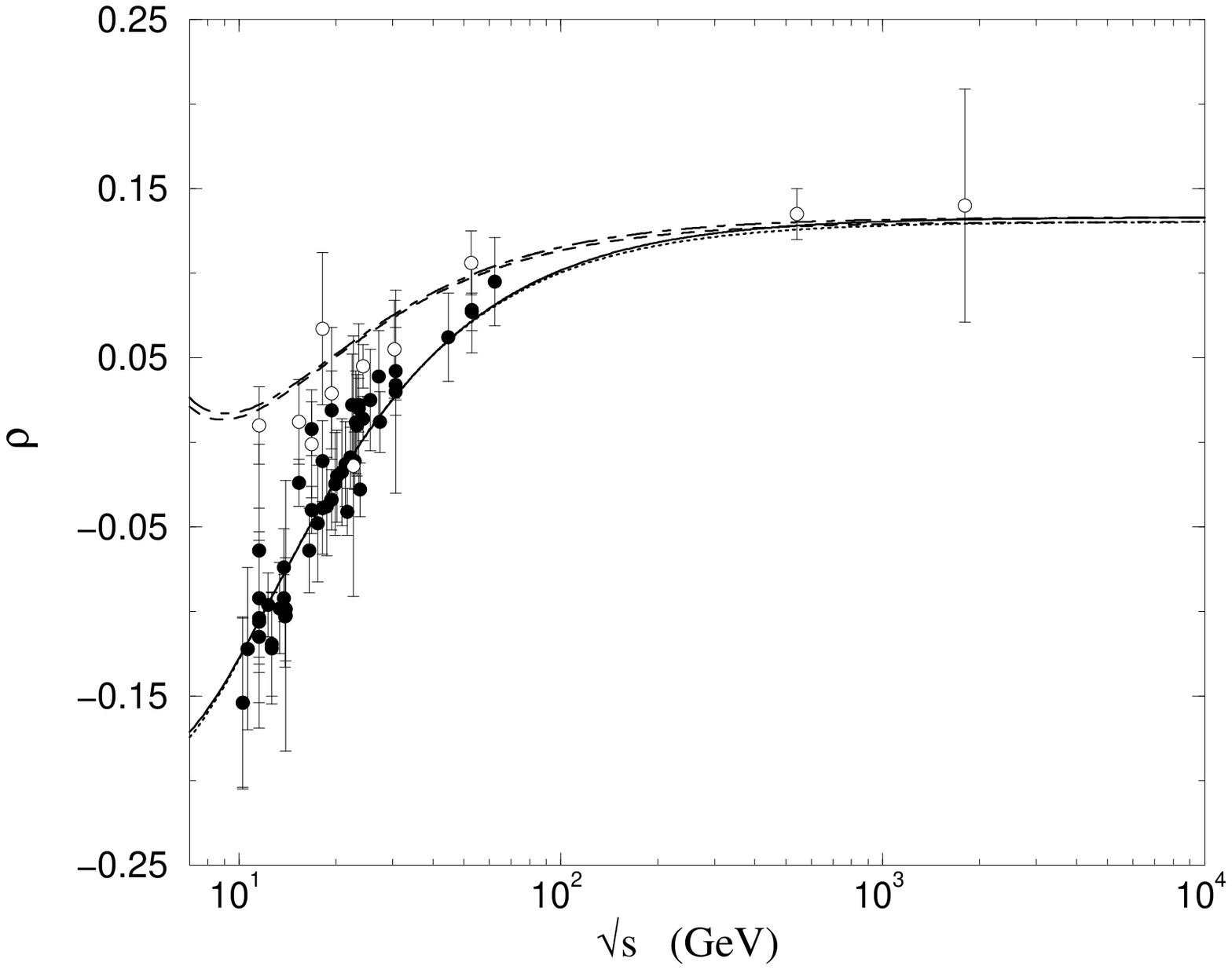,width=8.5cm,height=10cm}
\caption{\noindent
Simultaneous fits to $\sigma_{tot}(s)$ and $\rho(s)$ through the DL
parametrization with $K$ as free parameter and
ensembles I (dotted curves for $pp$ and dashed for $\bar{p}p$) and II 
(solid curves for $pp$ and dot-dashed for $\bar{p}p$)---columns 8 and 9 
in Table II.} 
%\label{rates}
\end{figure}

\clearpage

\newpage

\begin{table*}
\caption{Results with the Kang-Nicolescu model.} 
\label{kn}
\begin{ruledtabular}
\begin{tabular}{ccccccccc}
Fit: & \multicolumn{4}{c}{Individual } & \multicolumn{4}{c}{Global} 
\\ 
Quantity: & \multicolumn{2}{c}{\st} & \multicolumn{2}{c}{\ro} 
 & \multicolumn{4}{c}{\st and \ro}\\ 
Ensemble:  &  I &  II &  I &  II  &  I &  II &  I &  II\\ 
\hline
No. of points  &   $102$    &   $102$ & $63$ &$63$ & $165$ & $165$ & 
165& 165\\  
$\chi^2/\textrm{DOF}$ &0.78 &0.91 &1.05 &35.2 &0.78 &0.87 &0.77 
&0.86 \\  
$A_1$ (mb) & 44$\pm$1 & 47$\pm$1  & - & - & 44.7$\pm$0.6 & 
45.2$\pm$0.6 & 
44.4$\pm$0.6 & 45.0$\pm$0.7  \\   
$A_2$ (mb) & 45$\pm$3 & 52$\pm$3 & - & - & 44.7$\pm$0.7 & 
45.2$\pm$0.7 & 
44.5$\pm$0.7 & 45.1$\pm$0.7 \\  
$B_1$ (mb) & -2.9$\pm$0.3 & -3.6$\pm$0.4 & - & - & -3.0$\pm$0.2 
& -3.1$\pm$0.2 
& -2.9$\pm$0.2 & -3.1$\pm$0.2 \\ 
$B_2 (mb)$  & -2.9$\pm$0.6 &  -4.2$\pm$0.6 & - & - & -2.9$\pm$0.2
 & -3.1$\pm$0.2 
& -2.9$\pm$0.2 & -3.1$\pm$0.2 \\  
$k$ (mb)& 0.33$\pm$0.03  &  0.39$\pm$0.03 & - & - & 0.34$\pm$0.01 &
 0.35$\pm$0.01 
& 0.33$\pm$0.01 & 0.35$\pm$0.01 \\  
$R$ (mb) & 24$\pm$7  &  12$\pm$7 & - & - & 25.9$\pm$0.8 & 25.9$\pm$0.8 
& 25.3$\pm$0.9 & 25.4$\pm$0.9 \\ 
\hline
$K$ &  -  &  - & -198$\pm$32 & -1369$\pm$32 & 0 & 0 & -72$\pm$46 & 
-63$\pm$46 \\ 
\hline
Figure: & \multicolumn{2}{c} {5(a) and 5(b)} & \multicolumn{2}{c}{5(a)
 and 5(c)} 
 & \multicolumn{2}{c} {6} & \multicolumn{2}{c}{7}\\
\end{tabular}
\end{ruledtabular}
\end{table*}

\begin{figure}
\psfig{file=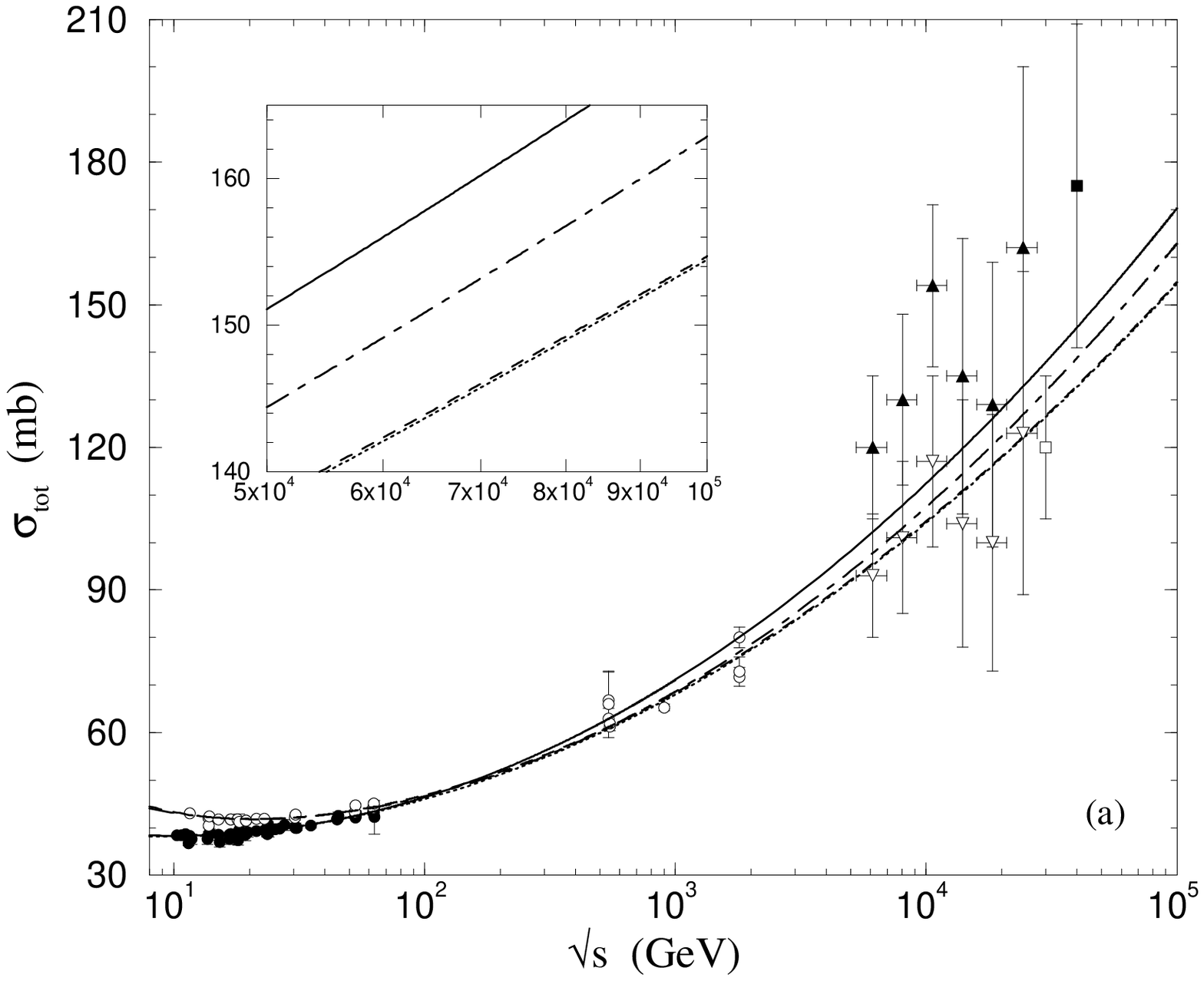,width=5.9cm,height=10cm}
\psfig{file=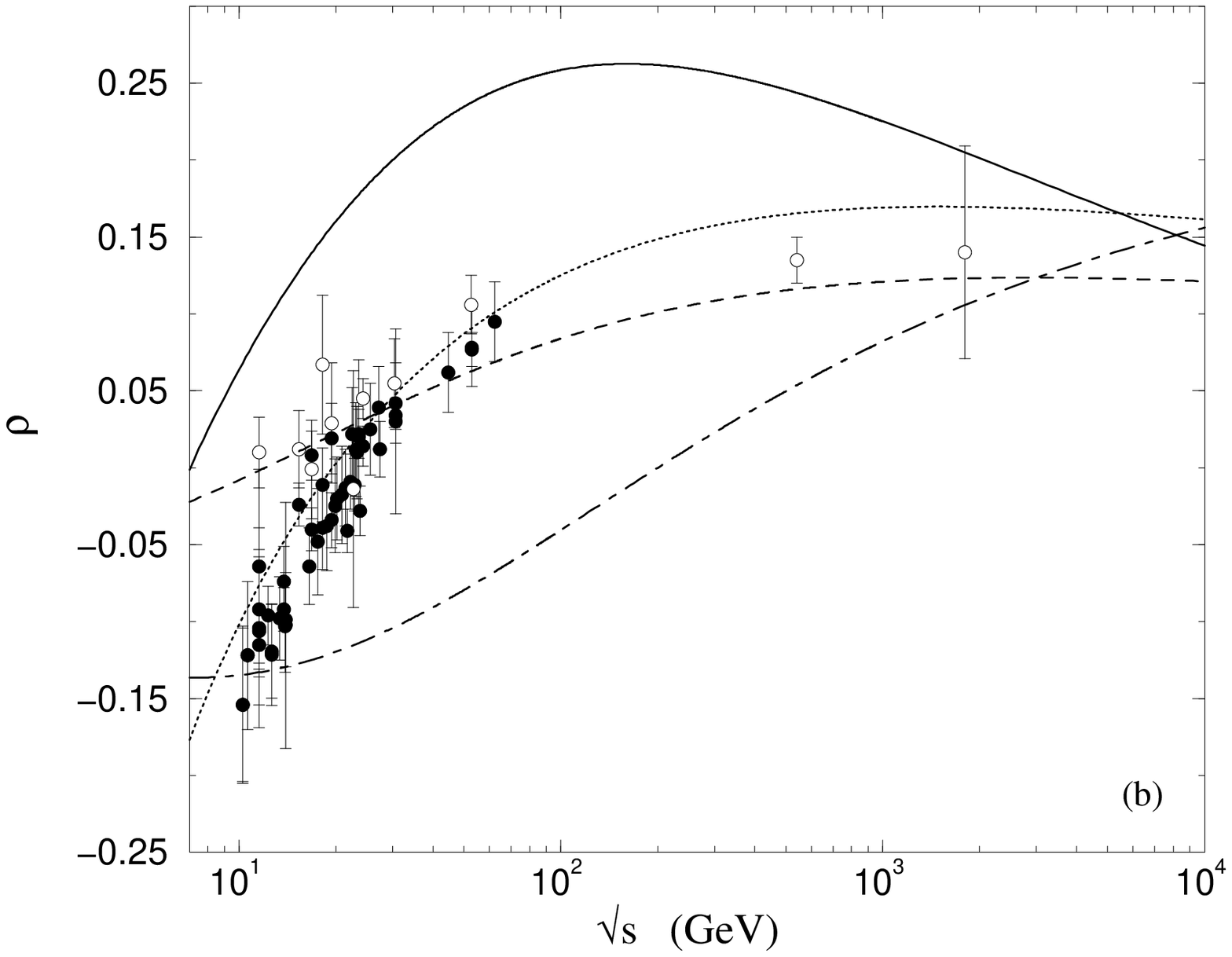,width=5.9cm,height=10cm}
\psfig{file=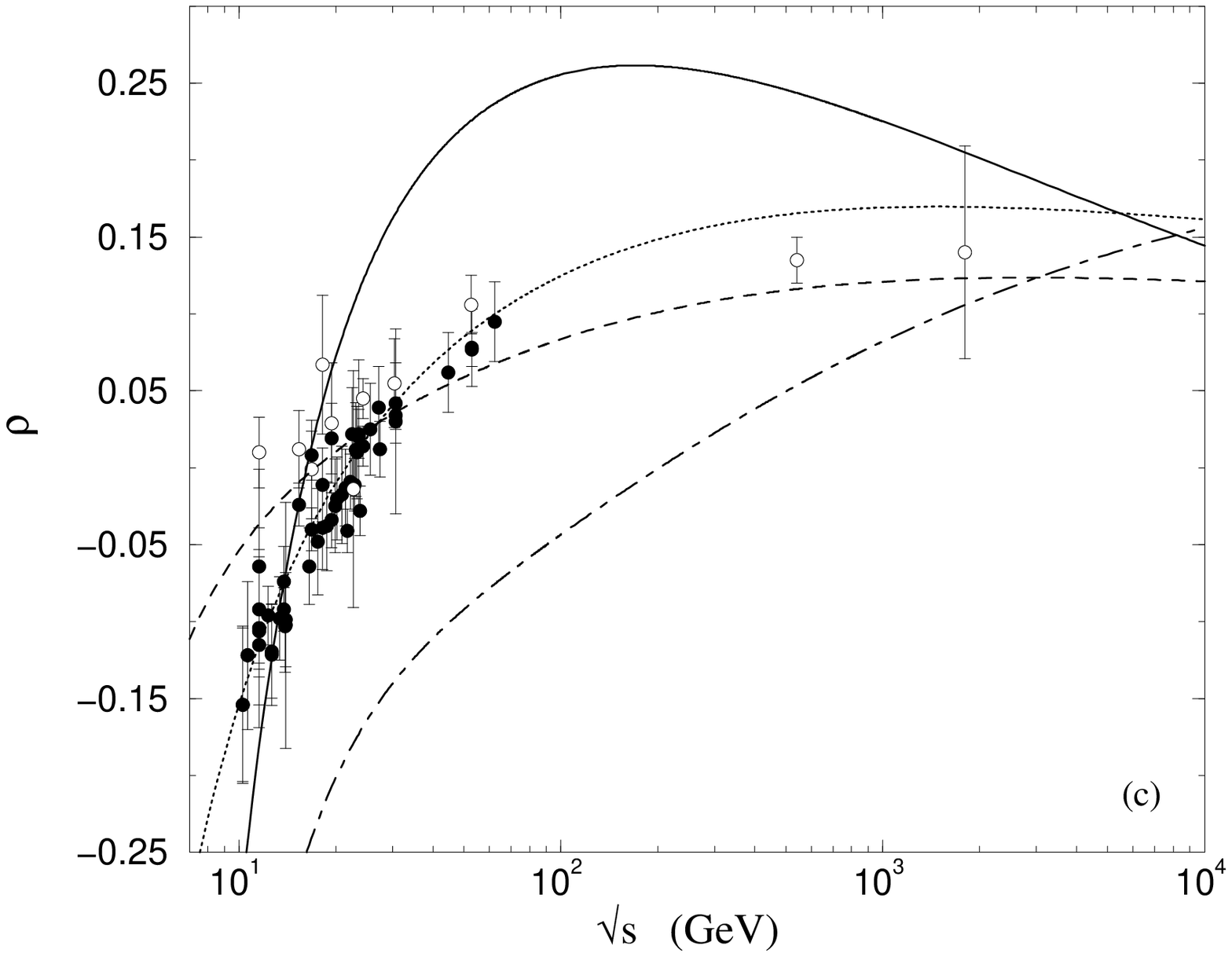,width=5.9cm,height=10cm}
\caption{\noindent
Fits to $pp$  and $\bar{p}p$  total cross section data from
ensembles I (dotted curves for $pp$ and dashed for $\bar{p}p$) and II 
(solid curves for $pp$ and dot-dashed for $\bar{p}p$), through the KN 
parametrization (a) and the
corresponding predictions for $\rho(s)$ with $K=0$ (b) and $K$ as free 
fit parameter
(c)---columns 2 and 3 in Table III.}
%\label{rates}
\end{figure}

\begin{figure}
\psfig{file=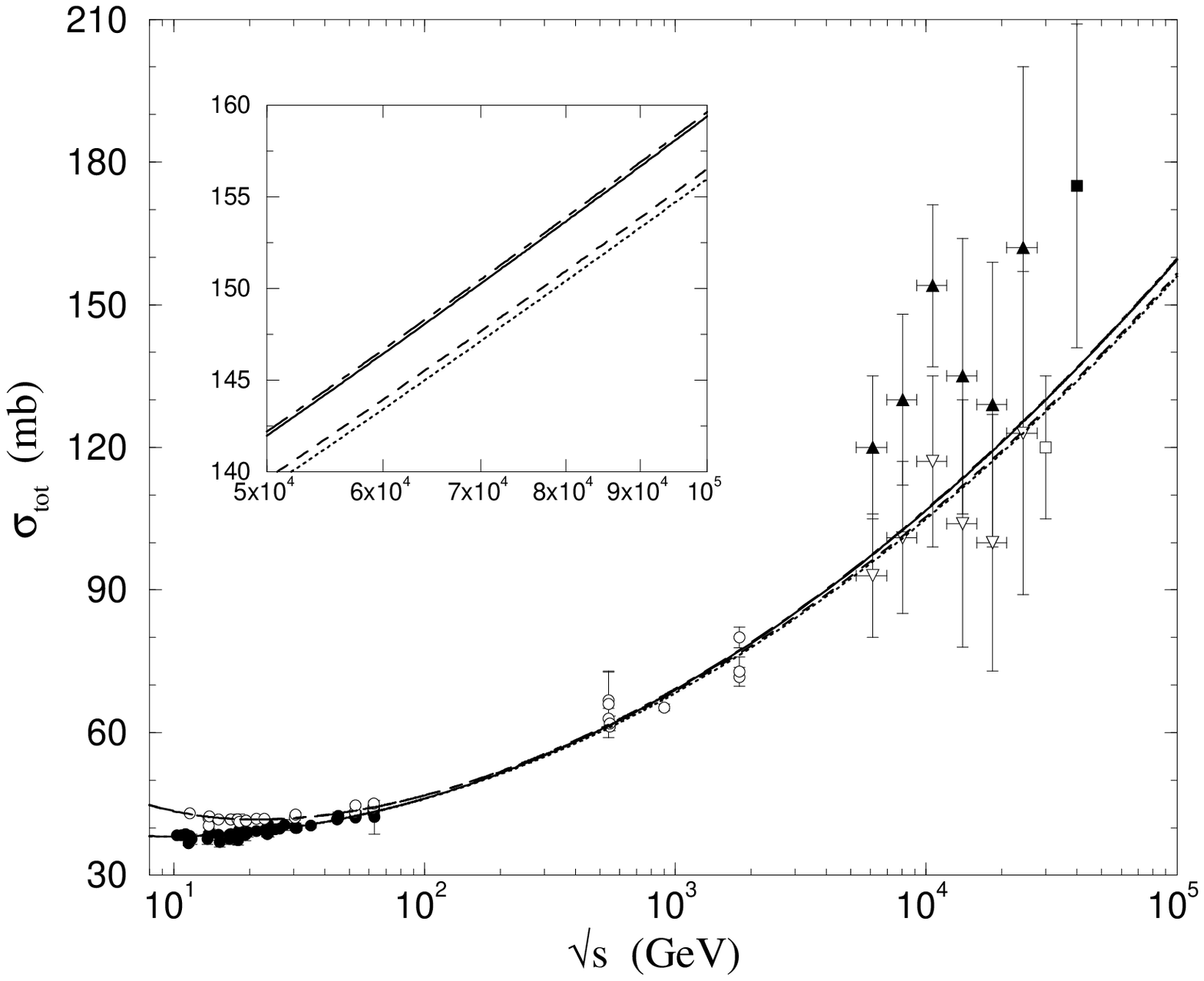,width=8.5cm,height=10cm}
\psfig{file=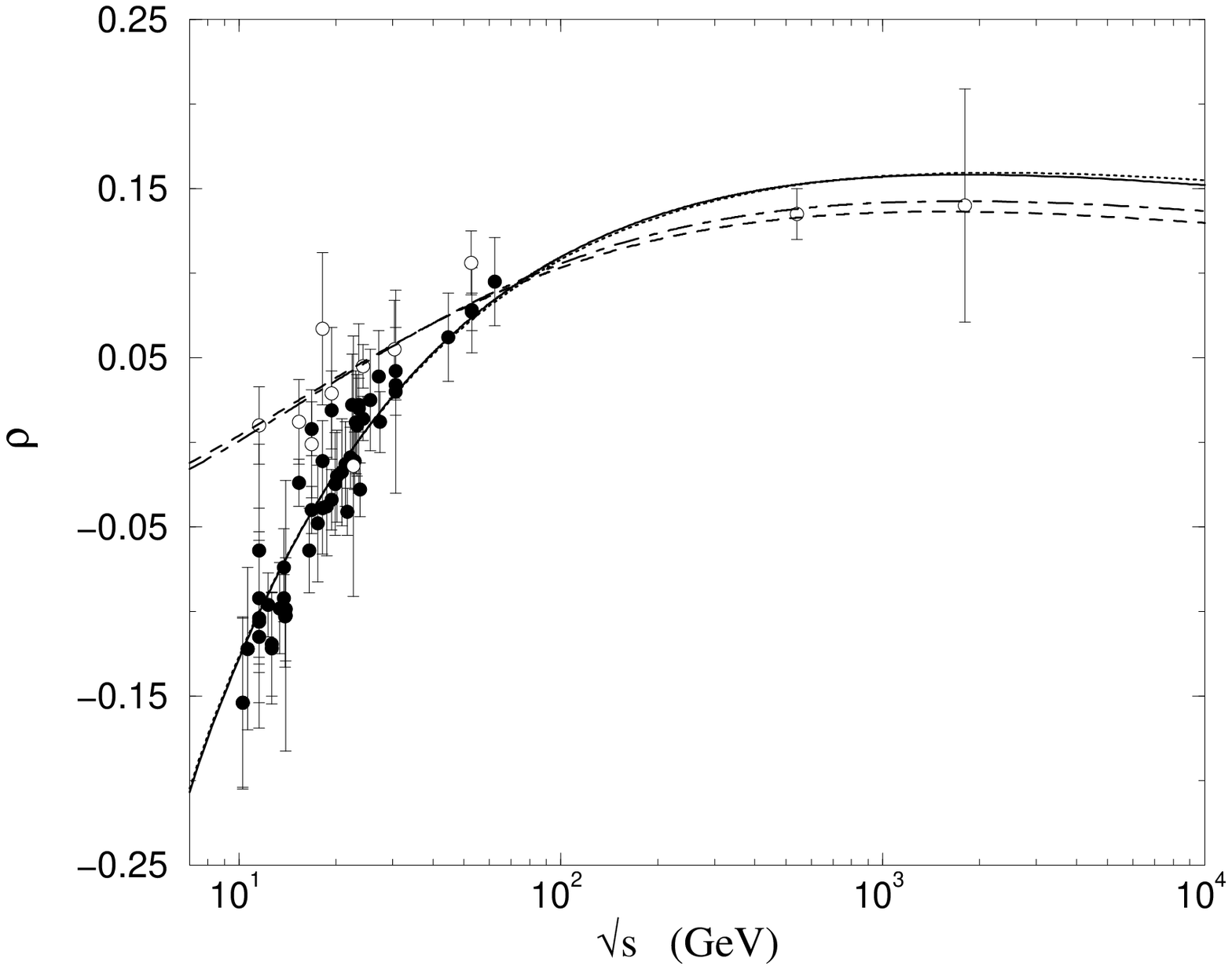,width=8.5cm,height=10cm}
\caption{\noindent
 Simultaneous fits to $\sigma_{tot}(s)$ and $\rho(s)$ through the KN
parametrization with $K=0$ and ensembles I (dotted curves for $pp$ and dashed 
for $\bar{p}p$) 
and II (solid curves for $pp$ and dot-dashed for $\bar{p}p$)---columns 6 and 7 
in Table III.}
%\label{rates}
\end{figure}

\begin{figure}
\psfig{file=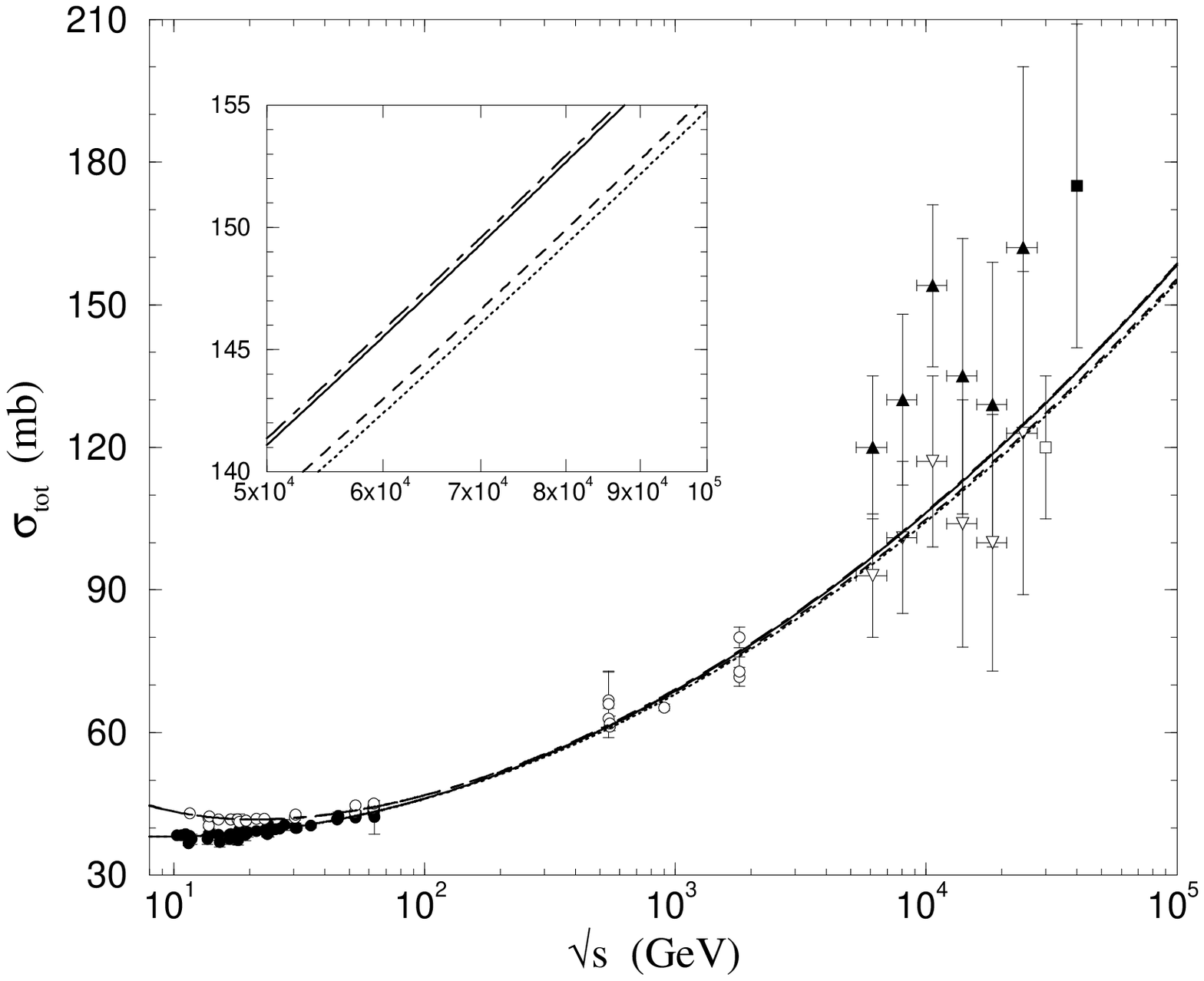,width=8.5cm,height=10cm}
\psfig{file=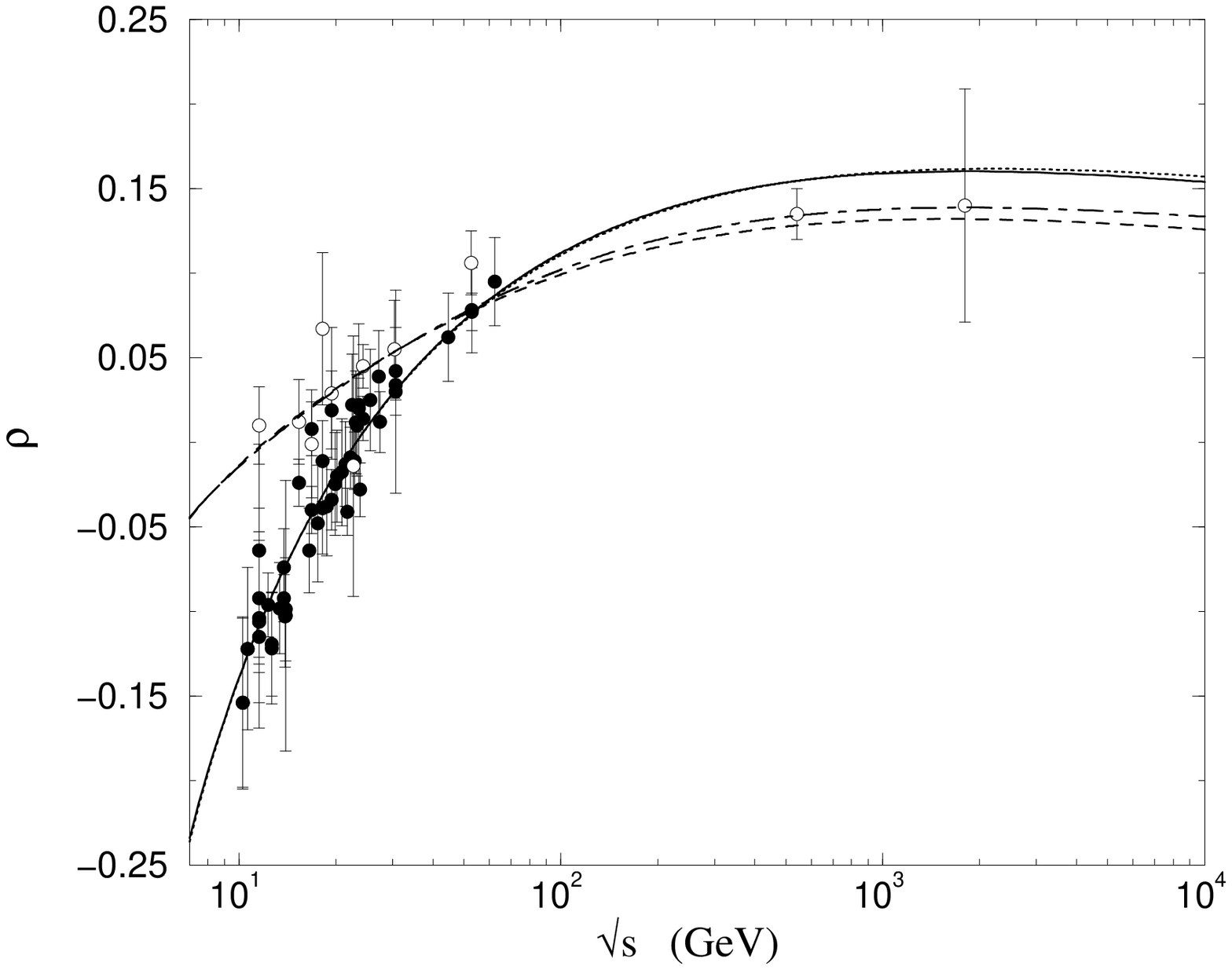,width=8.5cm,height=10cm}
\caption{\noindent
Simultaneous fits to $\sigma_{tot}(s)$ and $\rho(s)$ through the KN
parametrization with $K$ as free parameter and  ensembles I (dotted curves 
for $pp$ and dashed 
for $\bar{p}p$) and II 
(solid curves for $pp$ and dot-dashed for $\bar{p}p$)---columns 8 and 9 in 
Table III.} 
%\label{rates}
\end{figure}

\clearpage

\newpage

\section{Conclusions and final remarks}

We shall focus our conclusions and present some discussion on the
 following five points.

\vspace{0.3cm}

\textit{Ensembles I and II}.
The figures corresponding to fits to the total cross section data show 
that,
in general,  the results
with ensembles I and II do not differ substantially, except in the case 
of
individual fits with the KN model [Fig. 5.(a)]. Also, the cosmic-ray 
information in ensemble II (Nikolaev and GSY) is not described in all the 
cases.
This is a consequence of the small number of points and the large error 
bars
in comparison with the accelerator data, as well as the choices for the 
parametrizations 
(the models used). However, we must stress that, despite being a 
nonstandard result,
the cosmic-ray information in ensemble II 
has a reasonable basis,
as discussed in Sec. II B.
Certainly, more precise data are necessary for a truly conclusive result,
but, at present, it seems reasonable and interesting to investigate the 
semiquantitative
consequences of this nonstandard possibility.

Based on these ideas, in what follows, we shall consider the results
with both ensembles. However, it must also be noted that some of the 
conclusions that
follow are
independent of the ensemble used, such as those concerning  
global vs individual fits and
the role of the subtraction constant.

\vspace{0.3cm}

\textit{Bounds for the pomeron intercept}.
With the DL model and the cosmic-ray information in ensembles I and II,
 we may infer some novel limit 
values for the soft Pomeron intercept  $\alpha_{{\tt I\!P}}(0)= 1 + \epsilon$.
 From
Table II the highest (lowest) $\epsilon$ value was obtained with 
ensemble II (I), in
the case of individual fits to \st (global fits to \st and $\rho$), namely,

\vspace{0.2cm}

\centerline{$
\epsilon_{{\rm upper}} = 0.094 \qquad {\rm and} \qquad 
\epsilon_{{\rm lower}} = 0.079.
$}

\vspace{0.2cm}

\textit{Odderon}.
In the case of \textit{individual} fits with the KN model, Table III 
and Fig. 5
show that with ensemble II the model predicts a crossing in $\sigma_{tot}(s)$,
so that $\sigma_{tot}^{pp}$ becomes greater than 
$\sigma_{tot}^{\bar{p}p}$ above
$\sqrt s \approx 50$ GeV: $\Delta A = 5 \pm 3$ and $\Delta B \sim 0$ 
(Sec. II C). 
However, as shown
in Fig. 5, in this case the $\rho(s)$ data are not described 
($K = 0$ or $K$ as a free parameter).

On the other hand, Table III shows that, in the case of 
\textit{global fits} with both ensembles 
I and II, statistically, $\Delta A = 0$ and $\Delta B = 0$, so 
that $\Delta \sigma_{tot} =
\sigma_{tot}^{\bar{p}p} - \sigma_{tot}^{pp} = 0$. However, from Figs.
6 and 7, we see that, in both case, $K=0$ and $K$ as a free parameter, 
this model predicts
a crossing in the $\rho(s)$ behavior, with $\rho_{pp}$ becoming 
greater than
$\rho_{\bar{p}p}$ at $\sqrt s \approx 70-80$ GeV (for $K=0$) and 
$\sqrt s \approx 60-70$ GeV 
(for $K$ a free parameter), a result in agreement with the early 
fits by
Gauron, Nicolescu, and Leader \cite{gnl}. Most important, these 
results are 
also in complete agreement with
all the experimental data presently available on \st and \ro and 
are independent of
the ensemble used. This is certainly an
interesting and important prediction that will be verified at the 
BNL Relativistic Heavy Ion Collider
(RHIC) and the CERN Large Hadron Collider (LHC) \cite{exp}.

\vspace{0.3cm}

\textit{Individual and global fits}.
From Tables II and III, with the exception of the individual fit to 
ensemble II
with the KN model and $K$ as a free parameter, the statistical 
information does not 
indicate a preference between individual or global fits.

On the other hand, global fits clearly constrain the possible 
increase of the
total cross section. For example, in the case of the DL model, 
with both ensembles
I and II and $K = 0$, we obtained

\vspace{0.2cm}

\centerline{$
\epsilon_{\textrm{indi}}^{\textrm{I}} \approx 0.088  \quad \rightarrow
\quad \epsilon_{\textrm{simul}}^{\rm {I}} \approx 0.081 \quad 
(\textrm{reduction} \approx 9\%) 
$},

\vspace{0.2cm}

\centerline{$
\epsilon_{\textrm{indi}}^{\textrm{II}} \approx 0.091  \quad \rightarrow
\quad \epsilon_{\textrm{simul}}^{\textrm{II}} \approx 0.083 \quad 
(\textrm{reduction} \approx 10\%) \nonumber 
$}.

\vspace{0.2cm}

As discussed previously, \st and \ro do not have the same status 
as physical quantities, since 
the extracted \ro value is model dependent. 
Therefore, in principle, we understand that global fits underestimate 
the possible rise of \st
in the asymptotic region. In our analysis the effects of the global 
and individual
fits depend also on the subtraction constant, as discussed in what 
follows.

\vspace{0.3cm}

\textit{Subtraction constant}.
As commented before, if $K$ is taken as a free parameter, the fit
 procedure demands that it 
is correlated with all the other parameters of the model involved. 
Therefore, in
principle, it is expected to have effects in both the low- and 
high-energy regions. Let
us discuss our results through individual and global fits.

In the case of \textit{individual} fits with the DL model, Fig. 2 shows
 that the results 
for $\rho(s)$ with $K = 0$ and $K$ free are nearly the same above $\sqrt s 
\approx 40$ GeV
and that below this energy the predictions are quite different.
 We shall not discuss
the corresponding results with the KN model, since the data are not 
described (Fig. 5).

In the case of \textit{global} fits with the DL model, beyond the 
differences in $\rho(s)$,
here below $\sqrt s \approx 40$ GeV, the asymptotic values of 
$\sigma_{tot}(s)$ are
different, namely, from Table II,

\vspace{0.2cm}

\centerline{$
\epsilon_{K=0}^{\textrm{I}} = 0.081 \pm 0.002  \quad \textrm{and}
\quad \epsilon_{K\ \textrm{free}}^{\textrm{I}} = 0.083 \pm 0.02 \quad 
(\textrm{increase} \approx 2.5\%), 
$}

\vspace{0.2cm}

\centerline{$
\epsilon_{K=0}^{\textrm{II}} = 0.083 \pm 0.002  \quad \textrm{and}
\quad \epsilon_{K\ \textrm{free}}^{\textrm{II}} = 0.084 \pm 0.02 \quad 
(\textrm{increase} \approx 1.2\%). 
$}

\vspace{0.2cm}

\noindent
In particular, with ensemble I, for $pp$ scattering at $\sqrt s = 14$
 TeV the fits indicate
$\sigma_{tot}^{pp} \sim 101$ mb for $K=0$ and $\sigma_{tot}^{pp} 
\sim 104$ mb for $K$ 
as a free parameter.
On the other hand, for global fits,  the KN model is not so sensitive 
to the influence
of the subtraction constant at least for $\rho(s)$ above $\sqrt s \approx 
20$ GeV, as shown in Figs.
6 and 7.
Therefore, in general, the subtraction constant affects the fit results 
in both the
low- and high-energy regions. Since it is mathematically justified in 
order
to control the convergence of the integral (or derivative) dispersion
 relation,
we understand that the subtraction constant cannot be neglected in 
the analytical
approach.

\begin{acknowledgments}
We are thankful to FAPESP for financial support (Contracts No. 00/13490-6,
 00/00991-7,
and No. 00/04422-7). We are also thankful to Professor M. M. Block for sending
us the BHS results (Table I)
and to Dr. J. Montanha Neto for a critical reading of the manuscript.
\end{acknowledgments}

\end{document}